\documentclass{IEEEtran}
\usepackage{amssymb}
\usepackage{amsfonts}
\usepackage{amsmath}
\usepackage{algorithm}
\usepackage{algorithmic}
\usepackage{tikz}
\usepackage{graphicx,cite,multicol,multirow,diagbox,booktabs,array}
\usepackage{color}
\usetikzlibrary{shapes.geometric, arrows}
\ifCLASSINFOpdf
\else
\fi
	\usepackage[caption=false,font=footnotesize,labelfont=rm,textfont=rm]{subfig}
	\usepackage{stfloats}
	\usepackage{hyperref}
	\usepackage{bm}
	\usepackage{color}
	\usepackage{xcolor}
\begin{document}
\title{Secure Transmission for Cell-Free Symbiotic Radio Communications with Movable Antenna: Continuous and Discrete Positioning Designs}

\author{\IEEEauthorblockN{Bin Lyu,~\IEEEmembership{Senior Member,~IEEE,}~Jiayu Guan,~Meng Hua,~\IEEEmembership{Member,~IEEE,}~Changsheng You,~\IEEEmembership{Member,~IEEE,}\\~ Tianqi Mao,~\IEEEmembership{Member,~IEEE,}~and 
Abbas Jamalipour,~\IEEEmembership{Fellow,~IEEE} }

\thanks{
This article was presented in part at the WCSP 2024, Hefei, China, Oct. 24-26, 2024 \cite{conference}. \emph{(Corresponding author: Meng Hua.)}

\IEEEcompsocthanksitem Bin Lyu and Jiayu Guan are with the School of Communications and Information Engineering, Nanjing University of Posts and Telecommunications, Nanjing 210003, China.  Meng Hua is with the Department of Electrical and Electronic Engineering, Imperial College London, London SW7 2AZ, UK. Changsheng You is with the Department of Electrical and Electronic
Engineering, Southern University of Science and Technology, Shenzhen
518055, China. Tianqi Mao is  with Beijing Institute of Technology, Beijing 100081, China. Abbas Jamalipour is with the School of Electrical and Computer Engineering, University of Sydney, Sydney, NSW 2006, Australia.
}}
\maketitle
\begin{abstract}
In this paper, we study a movable antenna (MA) empowered secure transmission scheme for reconfigurable intelligent surface (RIS) aided  cell-free symbiotic radio (SR) system. Specifically, the MAs deployed at distributed access points (APs) work collaboratively with the RIS to establish high-quality propagation links for both primary and secondary transmissions, {as well as suppressing} the risk of eavesdropping on confidential primary information. We consider both continuous and discrete MA position cases and maximize the secrecy rate of primary transmission under the secondary transmission constraints, respectively. For the continuous position case, we propose a two-layer iterative optimization method based on differential evolution with one-in-one representation (DEO), to find a high-quality solution with relatively moderate computational complexity. For the discrete position case, we first extend the DEO based iterative framework by introducing the mapping and determination operations to handle the  characteristic of discrete MA positions. To further reduce the computational complexity, we then design an alternating optimization (AO) iterative framework to solve all variables within a single layer. In particular, we develop an efficient strategy to derive the sub-optimal solution for the discrete MA positions, superseding the DEO-based method. Numerical results validate  the effectiveness of the proposed MA empowered secure transmission scheme along with its optimization algorithms.

\end{abstract}


\IEEEpeerreviewmaketitle
\section{Introduction}
The exponential growth of wireless networks {has been} driving a dramatic rise in Internet of Things (IoT) deployments, creating an unprecedented demand for both spectrum bands and energy resources to sustain seamless connectivity \cite{LYC1}. For example, assigning exclusive spectrum to individual IoT devices would demand roughly 76 GHz of bandwidth. Even with the implementation of cognitive radio to enable spectrum sharing, the overall demand could still reach up to 19 GHz, which is still unbearable \cite{LR}. {Additionally, conventional IoT devices typically rely on radio frequency (RF) hardware to transmit information,} which incurs high costs and substantial energy demands, particularly given the massive scale of IoT devices. However, the scarcity of spectrum resources and the goal of reducing energy consumption pose significant obstacles to the widespread deployment of IoT devices, emphasizing the urgent need for innovative technologies.
	
    As a promising solution to the obstacles above, symbiotic radio (SR) \cite{LYC2} has emerged and attracted considerable academic interest. In a classic SR communication system,  secondary transmission is parasitic in  primary transmission by modulating and backscattering  the primary signal to  convey information, achieving the reuse of the spectrum band and energy source from  primary transmission, and thus avoiding the need of additional spectrum resources and reducing the system energy consumption. 
    Due to the superior performance  in spectral efficiency  and energy efficiency,  SR systems have been widely studied  \cite{YH,ZC,LX,ZQ}.  
In \cite{YH}, an SR system with multiple backscatter devices (BDs) was modeled and the system energy efficiency was further maximized. In \cite{ZC}, the finite blocklength channel codes were adopted to accurately characterize the achievable rate of secondary transmission.
    However, conventional BDs each equipped with only a single or few antennas yield unsatisfactory secondary transmission performance since these BDs always suffer from double path-attenuation. To handle this constraint, recent studies have integrated reconfigurable intelligent surface (RIS) \cite{HZ} into SR systems \cite{LX,ZQ} since  RIS  can intelligently adjusting the phase shifts of numerous elements to build high-quality backscattering links. For example, a pioneering work about RIS aided SR systems was presented in \cite{LX}, in which the bit error rate (BER) of secondary transmission was minimized. Unlink \cite{LX} considering a multiple-input single-output model,  \cite{ZQ} studied an RIS aided multiple-input multiple-output  SR system, for which the transmit power minimization problem was formulated to reduce the system energy consumption.

However, the existing works, e.g., \cite{YH,ZC,LX,ZQ}, exclusively investigated SR systems constrained by conventional cellular architectures, where inherent inter-cell interference obstructs reliable and high-quality communications. 
     To address this limitation,  cell-free networking architecture has been developed, in which distributed access points (APs) can collaboratively serve all users without strict cell boundaries, thus effectively mitigating the impact of inter-cell interference \cite{EB}. To harness the synergistic benefits of SR  and cell-free networking architecture, it is a natural evolution to merge them  \cite{ZD2,LF,QS,ZY1}. \cite{ZD2} is the first work to study cell-free SR systems, in which a two-phase uplink training scheme for channel estimation was first proposed and then the rate-region of primary and secondary transmissions was  characterized. Unlink utilizing the BD as a secondary transmitter in \cite{ZD2}, \cite{QS,ZY1,LF} adopted the RIS as an alternative. In particular,  \cite{QS} extended  \cite{LX} to the cell-free model and designed a two-layer optimization algorithm to minimize the BER of  secondary transmission. In \cite{ZY1}, several  RISs were deployed in cell-free SR systems to construct the secondary transmission from distributed APs to multiple users.
    In \cite{LF}, the system performance enhancement scheme based the estimated channel state information (CSI) was further explored.

    It is widely recognized that the broadcast characteristic inherent in wireless communications may introduce significant security vulnerabilities  \cite{ZCheng1,SHu}, which are  particularly pronounced in SR systems, as malicious eavesdroppers (Eves) may intercept confidential information intended to  primary  transmission and/or secondary transmission.
    To overcome this deficiency, physical-layer security has been applied in SR systems \cite{LiangSecure,CJ,LyuRobust,Huiming}.   In \cite{LiangSecure} and \cite{CJ}, secure transmission schemes were considered for BD enabled SR systems, while  \cite{LyuRobust} and \cite{Huiming} extended to study this problem for RIS aided SR systems. While \cite{LiangSecure,CJ,LyuRobust,Huiming} advance secure transmission schemes for SR systems, there persists a fundamental limitation that  all transceivers in  \cite{LiangSecure,CJ,LyuRobust,Huiming} are with fixed position antennas (FPAs), which restricts the exploitation of spatial channel variations, thereby constraining the effective utilization  of spatial degrees of freedom (DoFs) for enhancing  the security performance.

    Recently, movable antenna (MA) \cite{MPA,WMa} (also known as fluid antenna \cite{FASurvey}) has emerged as a promising way of exploiting the spatial DoFs, attracting significant research attention for its ability to proactively reconfigure wireless channels via spatially adaptable antenna positioning \cite{MPA,WMa,Zheng,Hu,Pengcheng,Globally,WQQ1,ZLJB}. \cite{MPA} and \cite{WMa} were among the first to systematically implement the MA technology in wireless communications and creatively establish a field-response model to model the  associated channels. In \cite{Zheng}, the MA  was applied to support high-quality  multi-user communications by exploiting the adjustment of  antenna position and rotation. In  \cite{Hu}, the MA technology was applied to achieve secure transmission by enhancing  the strength of the desired signal while suppressing that of the eavesdropping signal. 
    The MA technology was further utilized in a wireless powered mobile edge computing system \cite{Pengcheng}. 
 Although continuous MA positioning in \cite{MPA,WMa,Zheng,Hu,Pengcheng} enables substantial performance gains, practical electro-mechanical constraints limit motion control to discrete adjustments and the corresponding impact on system performance was presented in \cite{WQQ1,Globally,ZLJB}. 
    Given the superior performance of MA, integrating it into SR systems is a promising progression \cite{ZC1,ZHUma}. In \cite{ZC1}, the authors pioneered the application of MA in BD enabled SR systems, demonstrating that the MA achieves significantly higher beam gains compared to the FPA array. In \cite{ZHUma}, the authors utilized both MA and RIS in  SR systems and designed a robust beamforming scheme. Although \cite{ZC1} and \cite{ZHUma} have made  attempts to apply the MA in SR systems, to the best of our knowledge, there are not studies exploring the application of MA to enhance the security  performance of SR systems. Thus, the persistent deficiency in \cite{LiangSecure,CJ,LyuRobust,Huiming}  has not been adequately addressed, which motivates  this paper.

In this paper, we propose an MA empowered secure transmission scheme for an RIS aided cell-free SR system to counteract eavesdropping from  malicious Eves. In the considered system, distributed APs each equipped with MAs cooperatively transmit distinct  signals to  primary users (PUs), aiming to meet diverse primary transmission demands. The RIS serves as the secondary transmitter to achieve secondary transmission from itself to secondary users (SUs) by reflecting the primary signals. The MAs and RIS are jointly utilized  to establish robust communication conditions  for both primary and secondary transmissions, while mitigating eavesdropping on primary information by the Eves. For both continuous and discrete position cases,  we maximize the minimum secrecy rate for the primary transmission under  the quality of service (QoS) constraints on secondary transmission, respectively. Compared to \cite{LiangSecure,CJ,LyuRobust,Huiming}, this work differs in the following two aspects. First, we consider a cell-free networking architecture to build the collaboration among all distributed APs, which can avoid the  inter-cell interference and  enable more flexible designs of transmit beamforming.  Second, we deploy the MAs at all the APs to attain additional DoFs for improving the legitimate system performance, which introduces new challenges in optimizing the continuous and discrete MA positions. Unlink \cite{MPA,WMa,Zheng,Hu,Pengcheng}, the coexistence of primary and secondary transmissions in SR systems renders the performance optimization more complicated due to differentiated QoS requirements, since the conventional methods in \cite{MPA,WMa,Zheng,Hu,Pengcheng} cannot be applied to solve our problems straightforwardly. In contrast to \cite{ZC1} and \cite{ZHUma}, we consider the secure transmission  under the diverse primary transmission demands and investigate the performance optimization for both continuous and discrete MA position cases. The main contributions of this paper are as follows:

	\begin{itemize}
		\item{To our best knowledge, this is the first work  to explore an MA empowered secure transmission scheme for RIS aided cell-free SR systems.  The MAs are deployed at distributed APs to achieve three objectives. First, the MAs are utilized to construct high-quality propagation links for the primary transmission from the APs to multiple PUs. Second, the MAs collaborate with the RIS to enable secondary transmission from the RIS to the SUs, ensuring  satisfactory performance. Third, the MAs and the RIS are jointly adjusted to counter eavesdropping on confidential information in primary transmission.
        To evaluate the performance limit provided by the MAs and tackle their practical mobility constraints, we take into account both continuous and discrete  position cases.} 

		\item{For the continuous position case, we aim to maximize the minimum primary secrecy rate under the QoS constraints on secondary transmission. To deal with the non-convex optimization problem, we develop a  two-layer iterative  framework. In the inner layer, an alternating optimization (AO) algorithm incorporating successive convex approximation (SCA) and penalty methods is developed to  optimize the transmit beamforming vectors at all APs and RIS phase shift matrix. In the outer layer, {we apply the differential evolution with one-in-one representation (DEO) method} \cite{DEO} to find the optimal MA positions with relatively moderate computational complexity and fast convergence speed.}

        \item{For the discrete MA position case, we consider a similar optimization problem under the constraint of discrete MA positions  and propose two efficient methods to solve it. We first extend the  two-layer iterative  framework  by introducing  the mapping and determination operations for the DEO method,  attaining the discrete position selection matrices with satisfactory accuracy.  To further reduce the computational complexity, we then propose an AO iterative framework, within which  the transmit beamforming vectors, RIS phase shift matrix, and discrete position selection matrices are optimized iteratively in a single layer. Specifically,  we apply the SCA, Schur complement, penalty method, binary variable relaxation, the mapping and determination operations to find the sub-optimal solution for the discrete MA positions.}

        \item{ We validate  the effectiveness of the proposed MA empowered secure transmission scheme and the proposed optimization algorithms through numerical results, which yield the following observations. First,  our MA empowered secure transmission scheme can improve the secrecy rate of primary transmission by up to $12.4\%$ over the FPA scheme. Second, the continuous MA positioning results in better system performance than the discrete MA positioning. Third, the performance of the  proposed AO iterative framework is very close to that of the  DEO-based two-layer iterative framework. }

	\end{itemize}

	The rest of the paper is structured as follows. Section \ref{system model} describes the model of the  MA empowered RIS-assisted cell-free SR system. Sections \ref{two-layer} and \ref{discrete} investigate the secrecy rate maximization problems  for continuous and discrete position cases, respectively. Numerical results evaluating the performance of the proposed schemes and algorithms are presented in Section \ref{numerical}, followed by the conclusion in Section \ref{conclusion}.

	\section{System Model}
	\label{system model}
	\begin{figure}
		\centering
		\subfloat[]
        {
			\includegraphics[scale=0.45]{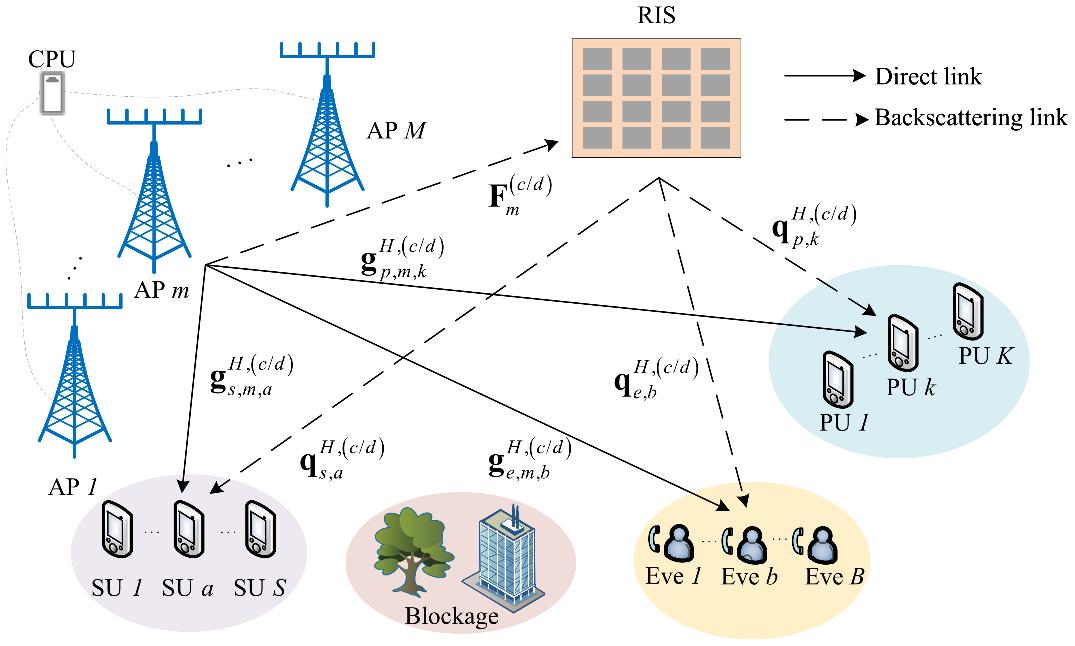}}\\
		\subfloat[]{
			\includegraphics[scale=0.9]{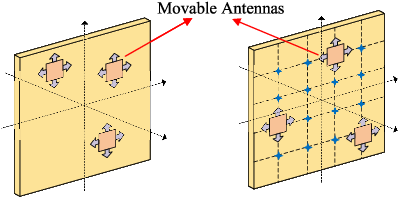}}
		\caption{{An MA empowered RIS-assisted cell-free SR system.
        (a): System model. (b): MAs for the continuous and discrete position cases, the marker $``$\textcolor{cyan}{$\blacklozenge$}$”$ represents discrete candidate positions.}
        }
		\label{fig:system_model}
	\end{figure}
	
	As illustrated in Fig.~\ref{fig:system_model}, we consider an MA empowered RIS aided cell-free SR system, which {is comprised of} $M$  distributed APs, $N_p$  PUs,  $N_s$  SUs, and one RIS. Each AP is equipped with $N$  MAs, the RIS has $N_\varepsilon$  reflecting {elements with fixed positions}, and the other devices are each {with a} single FPA. In addition, $N_e$  Eves {each equipped} with a single FPA are located near the PUs to intercept {confidential information} from the APs to the PUs. 
 Considering the presence of obstacles between the SUs and the Eves, we assume that eavesdropping on information at SUs is impossible \cite{LyuRobust}.
	In the considered system, the APs deliver distinct signals to all PUs. 
	The RIS is deployed to  reuse the spectrum and energy resources from  primary transmission to convey its information to the SUs. By employing flexible cabling to interconnect  RF chains and MAs at each AP, the MAs can adaptively reposition within a specified range  \cite{MPA,WMa}. 
    The collaborative design of the MA and RIS  constructs  optimized transmission links for primary and secondary transmissions.

	To simplify the subsequent description, let $\kappa = \{\varepsilon, p, e, s\}$  be the set composed of the RIS/PU/Eve/SU with $\varepsilon$, $p$, $e$ and $s$ being the symbols representing them, respectively. Similarly, $\xi = \{p\text{,}e\text{,}s\}$ is the set composed of the PU/Eve/SU.
		The sets corresponding to the MAs at each AP, the APs, the PUs, the Eves, the SUs, and the RIS elements are defined as $\mathcal{N}= \{1,\cdots, N\}$, $\mathcal{M}= \{1,\cdots, M\}$, $\mathcal{N}_p= \{1,\cdots, N_p\}$, $\mathcal{N}_e= \{1,\cdots, N_e\}$, $\mathcal{N}_s= \{1,\cdots, N_s\}$, and $\mathcal{N}_\varepsilon= \{1,\cdots, N_\varepsilon\}$, respectively.

	\subsection{Channel Model}
	Following \cite{MPA} and \cite{WMa}, we consider the far-field  response channel model, justified by the significant disparity in size between the transmit/receive region and the signal propagation distance.
Let $L_{t,\kappa}^m$ and $L_{r,\kappa}^m$ denote the number of transmit and receive channel paths between the $m$-th AP and node-$\kappa$, respectively, and the related sets of transmit and receive channel paths are defined as $\mathcal{L}_{t,\kappa}^m = \{1, \cdots, L_{t,\kappa}^m\}$ and $\mathcal{L}_{r,\kappa}^m = \{1, \cdots, L_{r,\kappa}^m\}$.
	For the $\varsigma$-th transmit path between the $m$-th AP and node-$\kappa$, the azimuth and elevation angles of departure (AoDs) are denoted as $\phi_{t,\varsigma}^{m,\kappa}$ and $\theta_{t,\varsigma}^{m,\kappa}$, respectively, where $\varsigma \in \mathcal{L}_{t,\kappa}^m$. 
	Similarly, $\phi_{r,\iota}^{m,\kappa}$ and $\theta_{r,\iota}^{m,\kappa}$ are the azimuth and elevation angles of arrival (AoAs) of the $\iota$-th receive path between the $m$-th AP and node-$\kappa$, {where $\iota\in \mathcal{L}_{r,\kappa}^m$. We consider the channel model with both continuous and discrete MA positions in the following.}
	
	\subsubsection{Continuous Position Case}
	\label{channel-con}
    For the continuous position case, the MAs are capable of unrestricted and continuous movement within predefined regions.
		The positions of the MAs at the $m$-th AP are denoted by $\bm{t}_m = [\bm{t}_{m,1}^T, \cdots, \bm{t}_{m,N}^T]^T$, with $\bm{t}_{m,n} = [x_{m,n}, y_{m,n}, z_{m,n}]^T$ representing the location of the $n$-th MA at the $m$-th AP in a three-dimensional local coordinate system, where $n \in \mathcal{N}$. 
        We define $\mathcal{C}_m$ as the movement region of the MAs at the $m$-th AP, assumed to be a cuboid with dimensions $A_1 \times A_2 \times A_3$. 
	Similarly, the local positions of the PUs, the Eves, the SUs, the RIS elements are denoted by $\bm{r}_p = [\bm{r}_{p,1}^T, \dots, \bm{r}_{p,N_p}^T]^T$, $\bm{r}_e = [\bm{r}_{e,1}^T, \dots, \bm{r}_{e,N_e}^T]^T$, $\bm{r}_s = [\bm{r}_{s,1}^T, \dots, \bm{r}_{s,N_s}^T]^T$, and $\bm{r}_{\varepsilon} = [\bm{r}_{\varepsilon,1}^T, \dots, \bm{r}_{\varepsilon,N\varepsilon}^T]^T$, respectively, where $\bm{r}_{\xi,\eta_{\xi}} = [x_{\xi,\eta_{\xi}}, y_{\xi,\eta_{\xi}}, z_{\xi,\eta_{\xi}}]^T$ is the coordinate of the $\eta_{\xi}$-th FPA at node $\xi$ in its local coordinate system, 
    $\bm{r}_{\varepsilon,\eta_{\varepsilon}} = [x_{\varepsilon,\eta_{\varepsilon}}, y_{\varepsilon,\eta_{\varepsilon}}, z_{\varepsilon,\eta_{\varepsilon}}]^T$ is the local coordinate of the $\eta_{\varepsilon}$-th element of the RIS 
    $\eta_{\xi} \in \mathcal{N}_\xi$, and $\eta_{\varepsilon} \in \mathcal{N}_\varepsilon$.

    For the link between the $m$-th AP and node-$\kappa$, the propagation difference of the signal between the position $\bm{t}_{m,n}$ and the transmit region origin for the $\varsigma$-th path is expressed as 
	\begin{align}
		\label{rhot}
	\rho_{t,\varsigma}^{m,\kappa}(\bm{t}_{m,n}) &= x_{m,n} \cos\theta_{t,\varsigma}^{m,\kappa} \cos\phi_{t,\varsigma}^{m,\kappa} \notag\\&+ y_{m,n} \cos\theta_{t,\varsigma}^{m,\kappa}\sin\phi_{t,\varsigma}^{m,\kappa} + z_{m,n} \sin\theta_{t,\varsigma}^{m,\kappa}.
	\end{align}
	In a similar manner, the signal propagation difference related with the $\bm{r}_{\kappa,\eta_\kappa}$ for the $\iota$-th receive path is given by
	\begin{align}
		\label{rhor}
	\rho_{r,\iota}^{m,\kappa}(\bm{r}_{\kappa,\eta_\kappa}) &= x_{\kappa,\eta_\kappa} \cos\theta_{r,\iota}^{m,\kappa} \cos\phi_{r,\iota}^{m,\kappa}\notag\\&+ y_{\kappa,\eta_\kappa} \cos\theta_{r,\iota}^{m,\kappa} \sin\phi_{r,\iota}^{m,\kappa}  + z_{\kappa,\eta_\kappa} \sin\theta_{r,\iota}^{m,\kappa},
	\end{align}
	where $\eta_\kappa \in\mathcal{N}_\kappa$ and $\kappa = \{\varepsilon, p, e, s\}$ represents the node type.\textbf{}
	
	Therefore, the channel between the $m$-th AP and the RIS can be characterized as \cite{MPA}, \cite{WMa}
	\begin{align}
		\label{Fm}
		{\bm{F}_{m}^{(c)}} = \bm{U}_{m,\varepsilon}^H \bm{\Sigma}_{m,\varepsilon} \bm{V}_{m,\varepsilon}(\bm{t}_m)\in \mathbb{C}^{N_\varepsilon \times N},
	\end{align}
	where $
	\bm{U}_{m,\varepsilon}(\bm{r}_\varepsilon) = [\bm{f}_{m,\varepsilon,1}(\bm{r}_{\varepsilon,1}), \cdots, \bm{f}_{m,\varepsilon,N_\varepsilon}(\bm{r}_{\varepsilon,N_\varepsilon})]\in\mathbb{C}^{L_{r,\varepsilon}^m \times N_\varepsilon}$ is the receive field-response matrix (FRM) of the link between the $m$-th AP and the RIS, $	\bm{f}_{m,\varepsilon,l}(\bm{r}_{\varepsilon,l}) = [e^{j \frac{2\pi}{\lambda}\rho_{r,1}^{m,\varepsilon}(\bm{r}_{\varepsilon,l})},\cdots, e^{j \frac{2\pi}{\lambda}\rho_{r,L_{r,\varepsilon}^m}^{m,\varepsilon}(\bm{r}_{\varepsilon,l})} ]^T\in\mathbb{C}^{L_{r,\varepsilon}^m \times 1}$ represents the field response vector (FRV), $l\in \mathcal{N}_\varepsilon$, $\bm{\Sigma}_{m,\varepsilon} \in \mathbb{C}^{L_{r,\varepsilon}^m \times L_{t,\varepsilon}^m}$ is the path-response matrix, $\lambda$ is the carrier wavelength, $\bm{V}_{m,\varepsilon}(\bm{t}_m) = [\mathbf{v}_{m,\varepsilon,1}(\bm{t}_{m,1}), \cdots, \mathbf{v}_{m,\varepsilon,N}(\bm{t}_{m,N})]\in \mathbb{C}^{L_{t,\varepsilon}^m \times N}$ is the transmit FRM with respect to $\bm{t}_m$, and  $\mathbf{v}_{m,\varepsilon,n}(\bm{t}_{m,n}) = [e^{j \frac{2\pi}{\lambda}\rho_{t,1}^{m,\varepsilon}(\bm{t}_{m,n})}, \cdots, e^{j \frac{2\pi}{\lambda}\rho_{t,L_{t,\varepsilon}^m}^{m,\varepsilon}(\bm{t}_{m,n})} ]^T\in\mathbb{C}^{L_{t,\varepsilon}^m \times 1}$ is the transmit FRV, $n\in\mathcal{N}$.

	The channel from the $m$-th AP to the $k$-th PU, the $b$-th Eve, and {the $a$-th SU} are denoted by $\bm{g}_{m,p,k}^{H,(c)}$, $\bm{g}_{m,e,b}^{H,(c)}$, {and $\bm{g}_{m,s,a}^{H,(c)}$,} respectively, where $k\in \mathcal{N}_p$, $b\in \mathcal{N}_e$, {and $a\in \mathcal{N}_s$.} The general structure of these channels is given as follows:
		\begin{align} 
			\label{g}
			\bm{g}_{m,\xi,\zeta}^{(c),H} = \bm{1}_{m,\xi}^H \bm{\Sigma}_{m,\xi,\zeta} \bm{V}_{m,\xi}(\bm{t}_m)\in \mathbb{C}^{1 \times N},
		\end{align}
		where $\xi = \{p, e, s\}$ represents the node type, $\zeta \in\mathcal{N}_\xi$ indexes the corresponding receiver, $\bm{1}_{m,\xi}$ is an $L_{r,\xi}^{m}\times 1$ all-ones column vector, $\bm{\Sigma}_{m,\xi,\zeta} \in \mathbb{C}^{L_{r,\xi}^m \times L_{t,\xi}^m}$ is the path-response matrix, 
	$\bm{V}_{m,\xi}(\bm{t}_m) = [\mathbf{v}_{m,\xi,1}(\bm{t}_{m,1}), \cdots, \mathbf{v}_{m,\xi,N}(\bm{t}_{m,N})]\in \mathbb{C}^{L_{t,\xi}^m \times N}$ is the transmit FRM, $\mathbf{v}_{m,\xi,n}(\bm{t}_{m,n}) = [e^{j \frac{2\pi}{\lambda}\rho_{t,1}^{m,\xi}(\bm{t}_{m,n})}, \cdots, e^{j \frac{2\pi}{\lambda}\rho_{t,L_{t,\xi}^m}^{m,\xi}(\bm{t}_{m,n})} ]^T\in\mathbb{C}^{L_{t,\xi}^m \times 1}$ is the transmit FRV.
	{In addition,  for $k\in \mathcal{N}_p$, $b\in \mathcal{N}_e$,  {and $a \in \mathcal{N}_s$,} {the channels from the RIS to the $k$-th PU, the $b$-th Eve, and the $a$-th SU, denoted by $\bm{q}_{p,k}^H \in \mathbb{C}^{1 \times N_\varepsilon}$, $\bm{q}_{e,b}^H \in \mathbb{C}^{1 \times N_\varepsilon}$, and $\bm{q}_{s,a}^H \in \mathbb{C}^{1 \times N_\varepsilon}$, are modeled in a similar way as \eqref{g}, respectively.}}

	\subsubsection{Discrete Position Case}
	Given that the inherent mobility of the antennas and practical electro-mechanical devices are limited to providing horizontal and vertical movement with a fixed constant increment \cite{Globally}, the positions of MAs may exhibit a discrete nature \cite{WQQ1} and must be selected from a quantized discrete set of candidate positions.
		For the discrete position case, we first define the set of all possible discrete candidate positions for the MAs at the $m$-th AP as $\bm{\mathcal{P}}_{m} = \left\{\bm{p}_{m,1}, \cdots, \bm{p}_{m,Q}\right\}$, where $\bm{p}_{m,q} = [x_{m,q}, y_{m,q}, z_{m,q}]^T$ corresponds to the $q$-th discrete candidate position at the $m$-th AP, with $q \in \mathcal{Q} = \{1, \cdots, Q\}$, and $Q$ denotes the number of candidate positions for each AP.
	The effective channel matrix that characterizes the channel coefficients between the RIS and all discrete candidate positions of the MAs at the $m$-th AP is denoted by $\widehat{\bm{F}}_{m} = \bm{U}_{m,\varepsilon}^H \bm{\Sigma}_{m,\varepsilon} \bm{V}_{m,\varepsilon}(\bm{\mathcal{P}}_{m})\in \mathbb{C}^{N_\varepsilon \times Q}$. 
		{Then, the} effective channel vectors from the $m$-th AP to the $k$-th PU, the $b$-th Eve, and {the $a$-th SU} are expressed as $\widehat{\bm{g}}_{m,p,k}^H\in \mathbb{C}^{1 \times Q}$, $\widehat{\bm{g}}_{m,e,b}^H\in \mathbb{C}^{1 \times Q}$, and {$\widehat{\bm{g}}_{m,s,a}^H\in \mathbb{C}^{1 \times Q}$,} respectively, with their corresponding expressions as  $\widehat{\bm{g}}_{m,\xi,\zeta}^H=\bm{f}_{m,\xi,\zeta}^H \bm{\Sigma}_{m,\xi,\zeta} \bm{V}_{m,\xi}(\bm{\mathcal{P}}_{m})$, {where $k\in \mathcal{N}_p$, $b\in \mathcal{N}_e$, $a\in \mathcal{N}_s$, and $\zeta \in\mathcal{N}_\xi$.}
		To model the channels for the discrete position case, we introduce a position selection matrix for the MAs at the $m$-th AP, which is given by $\bm{C}_m = [\bm{c}_{m,1}, \cdots, \bm{c}_{m,N}] \in \mathbb{C}^{Q \times N}$, where $\bm{c}_{m,n} = [\bm{c}_{m,n,1}, \cdots, \bm{c}_{m,n,Q}]^T \in \mathbb{C}^{Q \times 1}$, and $\bm{c}_{m,n,q} \in \{0, 1\}$. Specifically, $\bm{c}_{m,n,q} = 1$ signifies that the $n$-th MA at the $m$-th AP is placed at the $q$-th discrete candidate position. 
		Based on the above definitions, the channel model for all links associated with the APs for the discrete position case can be 	formulated as 
   \begin{align}
		\label{Fmdis}
		&\bm{F}_{m}^{(d)}=\widehat{\bm{F}}_{m} \bm{C}_m\in \mathbb{C}^{N_\varepsilon \times N},\\
		&\bm{g}_{m,\xi,\zeta}^{(d),H} = \widehat{\bm{g}}_{m,\xi,\zeta}^H\bm{C}_m\in \mathbb{C}^{1 \times N}.
	\end{align}The other channels unrelated with MAs are modeled in the same manner as in Section \ref{channel-con}.\footnote{ We consider a scenario where both PUs and Eves are  components of a primary communication system. Nevertheless, owing to Trojan virus infections, the Eves  attempt interception of conditional information transmitted to the PUs. Based on this assumption, the CSI associated with all devices can be   acquired by the channel estimation scheme proposed in \cite{HZ} and \cite{CEMA}, which guarantees the lower bound of secure transmission performance.}

	\subsection{Transmission Model}
	\label{trassmission}
  We adopt the PSR setup with the primary and secondary symbols sharing the same duration \cite{LR}, which applies to the scenarios such as smart cities and smart homes \cite{ZY1}.
	Let $s_k$ represent the primary symbol for the $k$-th PU, where $k \in \mathcal{N}_p$. The secondary symbol bearing specific information at the RIS is denoted as $c$. Both $s_k$ and $c$ follow an independent and identically distributed (i.i.d.) circularly symmetric complex Gaussian (CSCG) distribution, i.e., $s_k\sim\mathcal{CN}(0,1)$ and $c\sim\mathcal{CN}(0,1)$ \cite{LR}, \cite{ZY1}. The signal transmitted by the $m$-th AP is expressed as
		{$\bm{x}_m = \sum_{k=1}^{N_p}\bm{w}_{m,k}s_k$,} 
	where $\bm{w}_{m,k} \in \mathbb{C}^{N \times 1}$ denotes the transmit beamforming vector of the $m$-th AP  for the $k$-th PU. 
	{Furthermore, the phase shift matrix of the RIS is defined as $\bm{\Theta}=\text{diag}\{e^{j\theta_1},\cdots,e^{j\theta_{N_\varepsilon}}\}\in \mathbb{C}^{N_\varepsilon \times N_\varepsilon}$, where $\theta_l \in [0,2\pi)$ represents the phase shift of the $l$-th element, and $l \in \mathcal{N}_\varepsilon$.
We consider the scenario that all APs simultaneously send their independent primary signals to all receivers, including intended receivers (i.e., the PUs and SU) and unintended receivers (i.e., Eves).  
For $k \in \mathcal{N}_p$, $b \in \mathcal{N}_e$, and $a \in \mathcal{N}_s$, the received signals at the $k$-th PU, the $b$-th Eve, and the $a$-th SU, are denoted by $y_{p,k}^{(\imath)}$, $y_{e,b}^{(\imath)}$, and $y_{s,a}^{(\imath)}$, respectively, with the general expression as defined in \eqref{ypk2} \cite{ZY1}.
	\begin{align}
		\label{ypk2}
		y_{\xi,\zeta}^{(\imath)} 
		&= \sum\nolimits_{m=1}^M\bm{g}_{m,\xi,\zeta}^{(\imath),H} \bm{x}_m + \sqrt{\alpha}\sum\nolimits_{m=1}^M \bm{q}_{\xi,\zeta}^H\bm{\Theta}\bm{F}_m^{(\imath)} \bm{x}_mc+n_\xi\notag\\
        &=\bm{g}_{\xi,\zeta}^{(\imath),H}\bm{w}_{k}s_{k}+\bm{g}_{\xi,\zeta}^{(\imath),H}\sum\nolimits_{k'\neq k}^{N_p}\bm{w}_{k'}s_{k'}\notag\\&+\sqrt{\alpha}\bm{h}_{\xi,\zeta}^{(\imath),H}\sum\nolimits_{k'=1}^{N_p}\bm{w}_{k'}s_{k'}c+n_\xi,\zeta \in\mathcal{N}_\xi,\xi= \{p,e,s\},
	\end{align}
    where $\bm{w}_k = [\bm{w}_{1,k}^T,\cdots,\bm{w}_{M,k}^T]^T\in \mathbb{C}^{MN \times1}$, 
		$\bm{h}_{\xi,\zeta}^{(\imath)}=[ \bm{h}_{1,\xi,\zeta}^{(\imath),H},\cdots,\bm{h}_{M,\xi,\zeta}^{(\imath),H} ]^H \in \mathbb{C}^{MN \times1}$, $\bm{h}_{m,\xi,\zeta}^{(\imath)}=\bm{F}_m^{(\imath),H} \bm{\Theta}^H\bm{q}_{\xi,\zeta}\in \mathbb{C}^{N \times 1}$, $\bm{g}_{\xi,\zeta}^{(\imath)}=[ \bm{g}_{1,\xi,\zeta}^{(\imath),H},\cdots,\bm{g}_{M,\xi,\zeta}^{(\imath),H} ]^H \in \mathbb{C}^{MN \times1}$, and $\imath=\{c,d\}$ indicates whether the positions of the MAs are continuous or discrete.   
$\alpha$ is the reflection coefficient of the RIS, 
and $n_\xi  \sim \mathcal{CN}(0\text{,}\sigma_{\xi}^2)$ is the additive white Gaussian noise (AWGN). 
		It is evident that the first term of \eqref{ypk2} captures the intended signal transmitted through direct links from all distributed APs. The second and third terms of \eqref{ypk2} are the inter-user interference (IUI) and backscattering signals. When decoding the desired signal $s_k$, the IUI and the backscattering signal are considered as interference \cite{ZY1}. Hence, the signal-to-interference-plus-noise ratio (SINR) of the $k$-th PU for decoding $s_k$ is  $\gamma_{p,k,k}^{(\imath)}$, given by 
        \begin{align}
        \gamma_{p,k,k}^{(\imath)} = \frac{|\bm{g}_{p,k}^{(\imath),H} \bm{w}_k|^2 }{\sum_{k'\neq k}^{N_p}|\bm{g}_{p,k}^{(\imath),H} \bm{w}_{k'}|^2+\alpha\sum_{k'=1}^{N_p}|\bm{h}_{p,k}^{(\imath),H} \bm{w}_{k'}|^2  + \sigma_{p}^2 }
        \end{align}

        We assume that all Eves operate independently with unlimited computational resources, allowing them to cancel all IUI before decoding $s_k$ \cite{SHu}. {Thus, there is no IUI term in the SINR of the $b$-th Eve for decoding $s_k$, i.e., 
        \begin{align}
          \gamma_{e,b,k}^{(\imath)} = \frac{|\bm{g}_{e,b}^{(\imath),H} \bm{w}_k|^2 }{\alpha\sum_{k'=1}^{N_p}|\bm{h}_{e,b}^{(\imath),H} \bm{w}_{k'}|^2  + \sigma_{e}^2 }.
        \end{align}
		To enable the SUs to decode all $s_k$ and $c$ for $k\in\mathcal{N}_p$, we first assume that $s_k$ with higher indices {has} worse channel conditions, i.e., the channel condition of $s_{k_2}$ is inferior to that of $s_{k_1}$ for $k_1< k_2$, $k_1, k_2\in\mathcal{N}_p$ {\cite{MZsum}. Then, we apply the successive interference cancellation (SIC) technology to decode the $s_k$ with the best channel condition in the current round and remove it from \eqref{ypk2} before decoding the {$\hat{s}_k$} with the best channel condition in the next round \cite{MZsum}. This process continues until all $s_k$ are decoded, after which the SUs can decode $c$. Therefore, the SINR of {the $a$-th SU} for decoding $s_k$ {is $\gamma_{s,a,k}^{(\imath)}$, expressed as}
	\begin{equation}
		\label{rsk}
		\left \{
		\begin{aligned}
			&\frac{|\bm{g}_{s,a}^{(\imath),H} \bm{w}_k|^2 }{\sum_{k'=k+1}^{N_p}|\bm{g}_{s,a}^{(\imath),H} \bm{w}_{k'}|^2+\alpha\sum_{k'=1}^{N_p}|\bm{h}_{s,a}^{(\imath),H} \bm{w}_{k'}|^2  + \sigma_{s}^2 },\\ &~~~~~~~~~~~~~~~~~~~~~~~~~~~k\in\{1,\cdots,N_p-1\}, a\in\mathcal{N}_s,\\
			&\frac{|\bm{g}_{s,a}^{(\imath),H} \bm{w}_{N_p}|^2 }{\alpha\sum_{k'=1}^{N_p}|\bm{h}_{s,a}^{(\imath),H} \bm{w}_{k'}|^2  + \sigma_{s}^2 },a\in\mathcal{N}_s.
		\end{aligned}
		\right.
	\end{equation} Based on the above analysis, the achievable rates for the $k$-th PU, the $b$-th Eve, and the $a$-th SU of decoding $s_k$ are denoted by $R_{p,k,k}^{(\imath)}$, $R_{e,b,k}^{(\imath)}$, {and $R_{s,a,k}^{(\imath)}$,} respectively, expressed as 
    \begin{align}
      	R_{\xi,\zeta,k}^{(\imath)} = \log_2(1+\gamma_{\xi,\zeta,k}^{(\imath)}),~  k \in  \mathcal{N}_p, \zeta\in\mathcal{N}_\xi.  
    \end{align}

	For the $a$-th SU, after {all $s_k$} are decoded, the first term in \eqref{ypk2} can be eliminated, resulting in {the following expression}
		\begin{align}
			\label{ys'}			\hat{y}_{s,a}^{(\imath)}=\sqrt{\alpha}\bm{h}_{s,a}^{(\imath),H}\sum\nolimits_{k'=1}^{N_p}\bm{w}_{k'}s_{k'}c+n_{s}.
	\end{align}
According to  \cite{ZD2},     $\Xi_{s,a,k'}^{(\imath)} \triangleq  \sqrt{\alpha}\bm{h}_{s,a}^{(\imath),H}\sum\nolimits_{k'=1}^{N_p}\bm{w}_{k'}s_{k'}$ in \eqref{ys'} can be seen as a fast-fading channel for transmitting  $c$, and satisfies  $\Xi_{s,a,k'}^{(\imath)}\sim \mathcal{CN}(0\text{,}\delta_{s,a,k'}^{2,{(\imath)}})$, where $\delta_{s,a,k'}^{2,{(\imath)}}\triangleq\alpha\sum_{k'=1}^{N_p}|\bm{h}_{s,a}^{(\imath),H}\bm{w}_{k'}|^{2}$.
	Thus, the ergodic rate at {the $a$-th SU} {for decoding} $c$ can be formulated as \cite{LR}
	\begin{align}
			R_{c,a}^{(\imath)} &= \mathbb{E}_{\Xi_{s,a,k'}^{(\imath)}}[\log_2(1+{|\Xi_{s,a,k'}^{(\imath)}|^2}/{\sigma_{s}^2}) ]\notag\\&=-e^{{1}/{\gamma_{s,a,k'}^{(\imath)}}}\text{Ei}({-1}/{\gamma_{s,a,k'}^{(\imath)}})\log_2e,
	\end{align}
    where {$\text{Ei}({-1}/{\gamma_{s,a,k'}^{(\imath)}})$} denotes the exponential integral, and { $\gamma_{s,a,k'}^{(\imath)}={\alpha\sum_{k'=1}^{N_p}|\bm{h}_{s,a}^{(\imath),H}\bm{w}_{k'}|^{2}}/{\sigma_{s}^2}$.} 
	
	Similar to \cite{ZCheng1} and \cite{SHu}, we define the secrecy rate of primary communication at the $k$-th PU under the eavesdropping of the $b$-th Eve as $R_{b,k}^{sec,{(\imath)}}$, expressed as 
    \begin{align}
      R_{b,k}^{sec,{(\imath)}}=\left[R_{p,k,k}^{(\imath)}-R_{e,b,k}^{(\imath)}\right]^+,  
    \end{align}
	where $[\hat{x}]^+=\text{max}\left\{ \hat{x}\text{,}0 \right\}$, $k\in\mathcal{N}_p$, and $b\in\mathcal{N}_e$. The minimum secrecy rate for the primary transmission is defined as
    \begin{align}
        	R^{{sec},(\imath)}=\min_{k \in \mathcal{N}_p, b \in \mathcal{N}_e} \left\{ R_{b,k}^{sec,{(\imath)}} \right\}.
    \end{align}
		
	\section{Problem Formulation and Proposed Algorithm for Continuous Position Case}
	\label{two-layer}
	In this section, we aim to maximize the minimum secrecy rate of the primary communication for the continuous position case. {This goal is achieved} by jointly optimizing 
	the transmit beamforming vectors at the APs (i.e., $\{\bm{w}_{m,k}\}$, {$m\in\mathcal{M}$, $k\in\mathcal{N}_p$}), the phase shift matrix of the RIS (i.e., $\bm{\Theta}$), and the positions of the MAs (i.e., $\{\bm{t}_{m,n}\}$, {$m\in\mathcal{M}$, $n\in\mathcal{N}$}).
	The optimization problem {for the continuous position case is} formulated as
	\begin{subequations}
		\label{P1}
		\begin{align}
			&\max_{\{\bm{w}_{m,k}\},\bm{\Theta},\{\bm{t}_{m,n}\}}~~~\min_{k \in \mathcal{N}_p, b \in \mathcal{N}_e}\{ {R_{b,k}^{sec,{(c)}}} \}\\
			\label{15b}
			&~~~{\text{s. t.}}~~{\sum\nolimits_{k=1}^{N_p}}||\bm{w}_{m,k} ||^2 \le P_{\text{max}},m\in\mathcal{M},\\
			\label{15c}
			&~~~~~~~~~{R_{s,a,k}^{(c)}} \geq R_{th1},k\in\mathcal{N}_p,a\in\mathcal{N}_s,\\
			\label{15d}
			&~~~~~~~~~{R_{c,a}^{(c)}} \geq R_{th2},a\in\mathcal{N}_s,\\
			\label{15e}
			&~~~~~~~~~0 \leq \theta_l < 2\pi, l\in\mathcal{N}_\varepsilon,\\
			\label{15f}
			&~~~~~~~~~\bm{t}_{m,n}\in\mathcal{C}_m,m\in\mathcal{M},n\in\mathcal{N},\\
			\label{15g}
			&~~~~~~~~~\|\bm{t}_{m,n_1}-\bm{t}_{m,n_2}\|_2\geqslant D, n_1,n_2\in\mathcal{N}, n_1\neq n_2,
		\end{align}
	\end{subequations}
	where $P_{\text{max}}$ denotes the maximum  power provided by each AP.
	\eqref{15b} represents the power constraint imposed on the $m$-th AP, \eqref{15c} and \eqref{15d} ensure that the achievable rates for decoding $s_k$ and $c$ at the SUs satisfy the predefined thresholds $R_{th1}$ and $R_{th2}$, respectively. \eqref{15e} defines the allowable range of phase shifts,  
	\eqref{15f} constrains the movement range of the MAs, and 
	{\eqref{15g} indicates that the distance between {any two MAs remains} no less than the minimum distance $D$, thereby preventing coupling and collisions}  \cite{Globally}.
	Due to the coupling between the optimization variables, the objective {function in \eqref{P1}} is highly non-convex. Additionally, considering the non-convex constraints \eqref{15c}, \eqref{15d}, and \eqref{15g}, {problem \eqref{P1}} exhibits significant non-convexity. 
	To solve this problem, we first introduce an auxiliary variable $\chi$, and {reformulate} problem \eqref{P1} as follows
	\begin{subequations}
		\label{P2}
		\begin{align}
			&\max_{\{\bm{w}_{m,k}\},\bm{\Theta},\{\bm{t}_{m,n}\},\chi}~~~\chi\\
			&~~~~~~~~{\text{s. t.}}~~\eqref{15b}-\eqref{15g},\\
			\label{16c}
			&~~~~~~~~~~~~~~{R_{p,k,k}^{(c)}-R_{e,b,k}^{(c)}}\geq\chi,k\in\mathcal{N}_p,b\in\mathcal{N}_e.
		\end{align}
	\end{subequations}
	It is worth noting that \eqref{P2} remains a computationally intractable non-convex optimization problem. Thus, we develop a DEO-based two-layer iterative framework to address this challenge, where the AO algorithm is applied in the inner layer to optimize the transmit beamforming vectors at the APs and the phase shift matrix of the RIS, and the positions of the MAs are optimized using the DEO algorithm \cite{DEO} in the outer layer.

	\subsection{The Inner Layer}
	\label{inner}
	In the inner layer, with fixed $\{\bm{t}_{m,n}\}$, the joint design of $\{\bm{w}_{m,k}\}$ and $\bm{\Theta}$ is conducted. Due to the coupling of $\{\bm{w}_{m,k}\}$ and $\bm{\Theta}$, the non-convexity still exists. Thus, we adopt the AO algorithm to iteratively optimize $\{\bm{w}_{m,k}\}$ and $\bm{\Theta}$ {by solving \eqref{P3-1}, which is given by}
		\begin{align}
        \label{P3-1}
			\max_{\{\bm{w}_{m,k}\},\bm{\Theta},\chi}~\chi
			~~{\text{s. t.}}~~\eqref{15b}-\eqref{15e},\eqref{16c}.
		\end{align}
	The specific process of solving \eqref{P3-1} using the AO algorithm is summarized in Algorithm \ref{Algorithm1} and the details are described in the following.
	\subsubsection{Transmit Beamforming Optimization}
	\label{optw}
	{Given} $\{\bm{t}_{m,n}\}$ and $\bm{\Theta}$, we solve \eqref{P3} to obtain the optimal $\{\bm{w}_{m,k}\}$, as shown in steps 3-8 of Algorithm \ref{Algorithm1}. 
		\begin{align}
        		\label{P3}
			\max_{\{\bm{w}_{m,k}\},\chi}~\chi
			~~~{\text{s. t.}}~~\eqref{15b}-\eqref{15d},\eqref{16c}.
		\end{align}
	To overcome the non-convexity of \eqref{P3}, we define $\bm{W}_k=\bm{w}_k\bm{w}_k^H\in\mathbb{C}^{MN \times MN}$, where $\bm{W}_k\succeq0$,  $\text{Rank}(\bm{W}_k)=1$, and $k\in\mathcal{N}_p$. We further define auxiliary variables { $\bm{G}_{\xi,\zeta}^{(c)}=\bm{g}_{\xi,\zeta}^{(c)}\bm{g}_{\xi,\zeta}^{(c),H}\in\mathbb{C}^{MN \times MN}$, $\bm{H}_{\xi,\zeta}^{(c)}=\bm{h}_{\xi,\zeta}^{(c)}\bm{h}_{\xi,\zeta}^{(c),H}\in\mathbb{C}^{MN \times MN}$,} $\zeta\in\mathcal{N}_\xi$, and 
	$\bm{\Omega}_m = \text{diag}
	\left[
	\begin{array}{c}
		\underbrace{0,\cdots,0}_{(m-1)N}, 
		\underbrace{1,\cdots,1}_{N} , 
		\underbrace{0,\cdots,0}_{(M-m)N} 
	\end{array}
	\right]\in\mathbb{C}^{MN\times MN}$, $m\in\mathcal{M}$. Then, \eqref{15b} and \eqref{15c} can be reformulated as
	\begin{align}
		\label{18}
		{\sum\nolimits_{k=1}^{N_p}} \text{Tr}(\bm{W}_k \bm{\Omega}_m) \leq P_{\text{max}}, \quad m \in \mathcal{M},
	\end{align}
		\begin{equation}
			\label{19}
			\left \{
			\begin{aligned}
				&\frac{\text{Tr}(\bm{G}_{s,a}^{(c)} \bm{W}_k)}{\sum_{k' = k + 1}^{N_p} \text{Tr}(\bm{G}_{s,a}^{(c)} \bm{W}_{k'}) + \alpha \sum_{k' = 1}^{N_p} \text{Tr}(\bm{H}_{s,a}^{(c)} \bm{W}_{k'}) + \sigma_{s}^{2}}\\
				&~~~~~~~~~~~~~~~\geq 2^{R_{{th1}}} - 1, k \in \{1, \dots, N_p - 1\}, a\in\mathcal{N}_s, \\
				&\frac{\text{Tr}(\bm{G}_{s,a}^{(c)} \bm{W}_{N_p})}{\alpha \sum_{k' = 1}^{N_p} \text{Tr}(\bm{H}_{s,a}^{(c)} \bm{W}_{k'}) + \sigma_{s}^{2}}\geq 2^{R_{{th1}}} - 1, a\in\mathcal{N}_s.
			\end{aligned}
			\right.
		\end{equation}
		Note that $R_{c,a}^{(c)}$ is a non-decreasing function of {	{$\gamma_{s,a,k'}^{(c)}$.} Define} $\beta^*$ as the optimal solution of {$R_{c,a}^{(c)}=R_{th2}$,} \eqref{15d} can be thus rewritten as \eqref{20}, which is given by 
	\begin{align}
		\label{20}
		{\alpha\sum\nolimits_{k'=1}^{N_p}}\text{Tr}(\bm{H}_{s,a}^{(c)}\bm{W}_{k'})\geq \beta^* {\sigma_s^2}, a\in\mathcal{N}_s.
	\end{align}
    Furthermore, {$R_{p,k,k}^{(c)}-R_{e,b,k}^{(c)}$ in \eqref{16c} can be} reformulated as $\psi_1-\psi_2-\psi_3+\psi_4$, where 
    \begin{align}
    &\psi_1=\log_{2}(\Upsilon_{p,k}+\alpha\mathrm{Tr}(\bm{H}_{p,k}^{(c)}\bm{W}_{k})+\mathrm{Tr}(\bm{G}_{p,k}^{(c)}\bm{W}_{k})),\notag\\ 
    &\psi_2=\log_{2}(\Upsilon_{p,k}+\alpha\mathrm{Tr}(\bm{H}_{p,k}^{(c)}\bm{W}_{k}),\notag\\ 
    &\psi_3=\log_{2}(\Upsilon_{e,b}+\alpha\mathrm{Tr}(\bm{H}_{e,b}^{(c)}\bm{W}_{k})+\mathrm{Tr}(\bm{G}_{e,b}^{(c)}\bm{W}_{k})),\notag\\  
    &\psi_4=\log_{2}(\Upsilon_{e,b}+\alpha\mathrm{Tr}(\bm{H}_{e,b}^{(c)}\bm{W}_{k}),\notag\\
    &\Upsilon_{p,k}=\sum\limits_{k'\neq k}^{N_p}\mathrm{Tr}(\bm{G}_{p,k}^{(c)}\bm{W}_{k'})+\alpha\sum_{k'\neq k}^{N_p}\mathrm{Tr}(\bm{H}_{p,k}^{(c)}\bm{W}_{k'})+\sigma_{p}^{2},~~~~~\notag\\
    &\Upsilon_{e,b}=\alpha\sum\nolimits_{k'\neq k}^{N_p}\mathrm{Tr}(\bm{H}_{e,b}^{(c)}\bm{W}_{k'})+\sigma_e^2.\notag
   \end{align}
	To address the non-convexity resulting from the difference of concave functions, the first-order Taylor approximation technique is employed to obtain the upper bounds for 
	$\psi_2$ and $\psi_3$ at any feasible point $\bm{W}_k^{(s)}$ during the $s$-th iteration of the SCA method, which are expressed as 
		\begin{align}
			\label{taylor2}
			\psi_2&\leq\bar{\psi_2}\triangleq \log_{2}(\Upsilon_{p,k}+\alpha\mathrm{Tr}(\bm{H}_{p,k}^{(c)}\bm{W}_{k}^{(s)}))+\notag\\&\frac{\alpha\mathrm{Tr}(\bm{H}_{p,k}^{(c)}(\bm{W}_{k}-\bm{W}_{k}^{(s)}))}{(\Upsilon_{p,k}+\alpha\mathrm{Tr}(\bm{H}_{p,k}^{(c)}\bm{W}_{k}^{(s)}))\ln2},\\
			\label{taylor3}
			\psi_3&\leq\bar{\psi_3}\triangleq \log_{2}(\Upsilon_{e,b}+\alpha\mathrm{Tr}(\bm{H}_{e,b}^{(c)}\bm{W}_{k}^{(s)})+\mathrm{Tr}(\bm{G}_{e,b}^{(c)}\bm{W}_{k}^{(s)}))+\notag\\&\frac{\alpha\mathrm{Tr}(\bm{H}_{e,b}^{(c)}(\bm{W}_{k}-\bm{W}_{k}^{(s)}))+\alpha\mathrm{Tr}(\bm{G}_{e,b}^{(c)}(\bm{W}_{k}-\bm{W}_{k}^{(s)}))}{(\Upsilon_{p,k}+\alpha\mathrm{Tr}(\bm{H}_{e,b}^{(c)}\bm{W}_{k}^{(s)})+\mathrm{Tr}(\bm{H}_{e,b}^{(c)}\bm{W}_{k}^{(s)}))\ln2}.
	\end{align}
{Then, problem \eqref{P3} can be reformulated as problem \eqref{P4-22}:}
	\begin{subequations}
		\label{P4-22}
		\begin{align}
			&\max_{\bm{W}_k,\chi}~~~\chi~~~~~\\
			&~~{\text{s. t.}}~~\eqref{18}-\eqref{20},~\bm{W}_k\succeq0,\\
			&~~~~~~~~\psi_1-\bar{\psi_2}-\bar{\psi_3}+\psi_4 \geq \chi,\\
			\label{22d}
			&~~~~~~~~\text{Rank}(\bm{W}_k)=1.
		\end{align}
	\end{subequations}
	The semidefinite relaxation (SDR) technique is applied to relax \eqref{22d}, allowing the {CVX tool \cite{CVX}} to solve the relaxed version of \eqref{P4-22} and derive the rank-one solution. 
	The optimal solution of \eqref{P4-22}, denoted by $\bm{W}_k^*$, can be  obtained by iteratively implementing the SCA method. 
	\subsubsection{{RIS Phase Shift Matrix} Optimization}
	\label{RISopt}
	In steps 9-14 of Algorithm \ref{Algorithm1}, we optimize $\bm{\Theta}$ based on the {given $\{\bm{t}_{m,n}\}$ and $\bm{W}_{k}^*$} obtained from Section \ref{optw}, which requires solving the following problem
		\begin{align}
        	\label{P5-23}
			\max_{\bm{\Theta},\chi}~~~\chi
			~~~~~{\text{s. t.}}~~\eqref{15e},\eqref{16c},\eqref{19},\eqref{20}.
		\end{align}
	To enhance computational tractability of \eqref{P5-23}, auxiliary variables are incorporated {into it.} Let { $\bm{F}^{(c)}=[\bm{F}_1^{(c)},\cdots,\bm{F}_M^{(c)}]\in\mathbb{C}^{N_\varepsilon\times MN}$, $\bm{\Psi}_{\xi,\zeta}^{(c)}=\text{diag}\{\bm{q}_{\xi,\zeta}^{H}\}\bm{F}^{(c)}\in\mathbb{C}^{N_\varepsilon\times MN}$.} Moreover, let $\bm{\varphi}=[e^{j\theta_1},\cdots,e^{j\theta_{N_\varepsilon}}]^H\in\mathbb{C}^{N_\varepsilon \times 1}$,  $\bm{\Phi}=\bm{\varphi}\bm{\varphi}^H\in\mathbb{C}^{N_\varepsilon\times N_\varepsilon}$, where $\zeta\in\mathcal{N}_\xi$, $\bm{\Phi}\succeq0$, $\text{Rank}(\bm{\Phi})=1$. 
	Through the introduction of auxiliary variables, {$\bm{H}_{\xi,\zeta}^{(c)}$ can be reformulated 
		as $\bm{H}_{\xi,\zeta}^{(c)}=\bm{\Psi}_{\xi,\zeta}^{(c),H}\bm{\Phi}\bm{\Psi}_{\xi,\zeta}^{(c)}$. Building upon this,}
	\eqref{15e}, \eqref{19}, and \eqref{20} can be equivalently expressed as 
	\begin{align}
		\label{24}
		\bm{\Phi}(l,l)=1, l \in \mathcal{N}_\varepsilon, 
	\end{align}
		\begin{equation}
			\label{25-1}
			\left \{
			\begin{aligned}
				&\frac{\text{Tr}(\bm{G}_{s,a}^{(c)} \bm{W}_k^*)}{\Pi_1 + \alpha \sum_{k' = 1}^{N_p} \text{Tr}(\bm{\Psi}_{s,a}^{(c),H}\bm{\Phi}\bm{\Psi}_{s,a}^{(c)}\bm{W}_{k'}^*) + \sigma_{s}^{2}}\geq 2^{R_{{th1}}} - 1,\\
				&~~~~~~~~~~~~~~~~~~~~~~~~~~~~~~~~~~~~~~~ k \in \{1, \dots, N_p - 1\}, \\
				&\frac{\text{Tr}(\bm{G}_{s,a}^{(c)} \bm{W}_{N_p}^*)}{\alpha \sum_{k' = 1}^{N_p} \text{Tr}(\bm{\Psi}_{s,a}^{(c),H}\bm{\Phi}\bm{\Psi}_{s,a}^{(c)} \bm{W}_{k'}^*) + \sigma_{s}^{2}}\geq 2^{R_{{th1}}} - 1, 
			\end{aligned}
			\right.
		\end{equation}
	\begin{align}
		\label{25}
		\alpha \sum\nolimits_{k' = 1}^{N_p} \text{Tr}(\bm{\Psi}_{s,a}^{(c),H}\bm{\Phi}\bm{\Psi}_{s,a}^{(c)}\bm{W}_{k'}^*)\geq \beta^* {\sigma_s^2}, a\in\mathcal{N}_s,
	\end{align}
		where $\Pi_1=\sum_{k' = k + 1}^{N_p} \text{Tr}(\bm{G}_{s,a}^{(c)} \bm{W}_{k'}^*)$ and $a\in\mathcal{N}_s$.
		Then, we proceed to handle the non-convex constraint \eqref{16c} {by employing the SCA technique to convert it into a convex form,} reformulating it as
	\begin{align}
		\label{26}
		\varrho_1-\bar{\varrho_2}-\bar{\varrho_3}+\varrho_4 \geq \chi,
	\end{align}
   where
       \begin{align} 
		&\varrho_1=\log_{2}(\Lambda_{p,k}+\alpha\sum\nolimits_{k'=1}^{N_p}\mathrm{Tr}(\bm{\Psi}_{p,k}^{(c),H}\bm{\Phi}\bm{\Psi}_{p,k}^{(c)}\bm{W}_{k'}^*)),\notag\\
		&\bar{\varrho_2}\triangleq \log_{2}(\Gamma_{p,k}+\alpha\sum\nolimits_{k'=1}^{N_p}\mathrm{Tr}(\bm{\Psi}_{p,k}^{(c),H}\bm{\Phi}^{(s)}\bm{\Psi}_{p,k}^{(c)}\bm{W}_{k'}^*))\notag\\&+\frac{\alpha\sum\nolimits_{k'=1}^{N_p}\mathrm{Tr}(\bm{\Psi}_{p,k}^{(c),H}(\bm{\Phi}-\bm{\Phi}^{(s)})\bm{\Psi}_{p,k}^{(c)}\bm{W}_{k'}^*)}{(\Gamma_{p,k}+\alpha\sum\nolimits_{k'=1}^{N_p}\mathrm{Tr}(\bm{\Psi}_{p,k}^{(c),H}\bm{\Phi}^{(s)}\bm{\Psi}_{p,k}^{(c)}\bm{W}_{k'}^*))\ln2},\notag\\ 
		&\bar{\varrho_3}\triangleq \log_{2}(\Lambda_{e,b}+\alpha\sum\nolimits_{k'=1}^{N_p}\mathrm{Tr}(\bm{\Psi}_{e,b}^{(c),H}\bm{\Phi}^{(s)}\bm{\Psi}_{e,b}^{(c)}\bm{W}_{k'}^*))\notag\\&+\frac{\alpha\sum\nolimits_{k'=1}^{N_p}\mathrm{Tr}(\bm{\Psi}_{e,b}^{(c),H}(\bm{\Phi}-\bm{\Phi}^{(s)})\bm{\Psi}_{e,b}^{(c)}\bm{W}_{k'}^*)}{(\Lambda_{e,b}+\alpha\sum_{k'=1}^{N_p}\mathrm{Tr}(\bm{\Psi}_{e,b}^{(c),H}\bm{\Phi}^{(s)}\bm{\Psi}_{e,b}^{(c)}\bm{W}_{k'}^*))\ln2},\notag\\
		&\varrho_4=\log_{2}(\sigma_{e}^{2}+\alpha\sum\nolimits_{k'=1}^{N_p}\mathrm{Tr}(\bm{\Psi}_{e,b}^{(c),H}\bm{\Phi}\bm{\Psi}_{e,b}^{(c)}\bm{W}_{k'}^*)),\notag\\
       &\Lambda_{p,k}=\sum\limits_{k'=1}^{N_p}\mathrm{Tr}(\bm{G}_{p,k}^{(c)}\bm{W}_{k'}^*)+\sigma_{p}^{2},\ 
       \Lambda_{e,b}=\mathrm{Tr}(\bm{G}_{e,b}^{(c)}\bm{W}_{k}^*)+\sigma_{e}^{2},\notag\\
       &\Gamma_{p,k}=\sum\nolimits_{k'\neq k}^{N_p}\mathrm{Tr}(\bm{G}_{p,k}^{(c)}\bm{W}_{k'}^*)+\sigma_{p}^{2},\notag 
    \end{align}  
        and $\bm{\Phi}^{(s)}$ denotes the feasible solution for $\bm{\Phi}$ in the $s$-th iteration. Thus, \eqref{P5-23} is recast as
	\begin{subequations}
		\label{P6-27}
		\begin{align}
			\max_{\bm{\Phi}\succeq0,\chi}~~~\chi
			~~~~~&\text{s. t.}~~\eqref{24}-\eqref{26}, \\
			\label{27d}
			&~~~~~~~\text{Rank}(\bm{\Phi})=1.
		\end{align}
	\end{subequations}
	{To properly handle \eqref{27d},} 
	the penalty method \cite{JN} is employed to obtain a high-precision rank-one solution. {Accordingly, constraint} \eqref{27d} admits the following equivalent formulation
	\begin{align}
		\label{pen}
		||\bm{\Phi}||_*-||\bm{\Phi}||_2=0,
	\end{align}
	where the nuclear norm $||\bm{\Phi}||_*$ and spectral norm $||\bm{\Phi}||_2$ of  $\bm{\Phi}$ satisfy $||\bm{\Phi}||_* = \sum_{{m'}} \bar{\sigma}_{m'}(\bm{\Phi})$ and $||\bm{\Phi}||_2 = \bar{\sigma}_1(\bm{\Phi})$,  $\bar{\sigma}_{m'}(\bm{\Phi})$ represents the $m'$-th largest singular value. We incorporate constraint \eqref{27d} into the objective function through a penalty term $\hbar(||\bm{\Phi}||_*-||\bm{\Phi}||_2)$, thereby obtaining the following problem
		\begin{align}
        		\label{P7-28}
			\max_{\bm{\Phi}\succeq0,\chi}~~~\chi-\hbar(||\bm{\Phi}||_*-||\bm{\Phi}||_2)
			~~~~~{\text{s. t.}}~~\eqref{24}-\eqref{26},
		\end{align}
	{where $\hbar>0$ is the weight of the penalty term, and the solution to \eqref{P7-28} is guaranteed to satisfy \eqref{pen} when $\hbar\to\infty$.} However, \eqref{P7-28} constitutes a non-convex optimization problem. To solve this problem, {we apply a method similar to the one used for handling \eqref{16c}, where a first-order Taylor approximation of the penalty term $\hbar(||\bm{\Phi}||_*-||\bm{\Phi}||_2)$ is computed at any feasible point $\bm{\Phi}^{(s)}$ during the $s$-th iteration,} yielding the following convex upper bound
		\begin{align}
			\label{pentaylor}
			||\bm{\Phi}||_*-||\bm{\Phi}||_2&\leq\ell(\bm{\Phi},\bm{\Phi}^{(s)})\triangleq ||\bm{\Phi}||_*-||\bm{\Phi}^{(s)}||_2\notag\\&-\text{Tr}[\bm{\varpi}(\bm{\Phi}^{(s)})  \bm{\varpi}(\bm{\Phi}^{(s)})^H(\bm{\Phi}-\bm{\Phi}^{(s)})],
		\end{align}
		where $\bm{\varpi}(\bm{\Phi}^{(s)})$ represents the eigenvector corresponding to the largest eigenvalue of $\bm{\Phi}^{(s)}$. {Furthermore,} \eqref{P7-28} is transformed into a tractable convex optimization problem 
		\begin{align}
        	\label{P8-32}
			\max_{\bm{\Phi},\chi}~~~\chi-\hbar\ell(\bm{\Phi},\bm{\Phi}^{(s)})
			~~~~~{\text{s. t.}}~~\eqref{24}-\eqref{26}.
		\end{align}
By iteratively solving problem \eqref{P8-32} using the CVX tool \cite{CVX}, we obtain its optimal solution $\bm{\Phi}^*$. Then, the phase shift matrix of the RIS (i.e., $\bm{\Theta}^*$) can be expressed as $\bm{\Theta}^*=\text{diag}\{(\bm{\varphi}^*)^H\}$,
	where $\bm{\varphi}^*$ is the eigenvector with respect to the largest eigenvalue of $\bm{\Phi}^*$.
In order to mitigate the impact of the initialization of $\hbar$ on the  convergence performance, we first set it to a small value and update it via $\hbar^{(s+1)}=\vartheta \hbar^{(s)}$, where $\vartheta>1$ represents a step size.

	\begin{algorithm}
		\caption{ The proposed AO algorithm for solving \eqref{P3-1}.}
		\label{Algorithm1}
		\begin{algorithmic}[1]  
			\STATE {Initialize $\{ \bm{t}_{m,n}\}$, {$\bm{W}^{(\nu)}$, $\bm{\Theta}^{(\nu)}$,} the convergence threshold $\epsilon$ for the AO framework, the convergence threshold $\epsilon_1$ for the SCA method, and set {the initial} AO iteration index $\nu=0$.}
			\REPEAT
			\STATE{Initialize $\bm{W}^{(s)}$ {and set} the SCA iteration index $s=0$.}
			\REPEAT
			\STATE{Update $\bm{W}^{(s+1)}$ by solving \eqref{P4-22}.}
			\STATE{$s=s+1$.}
			\UNTIL{the increases of the objective function value of \eqref{P4-22} is less than $\epsilon_1$.}
			\STATE{Obtain $\bm{W}^*$ and let $\bm{W}^{(\nu)}=\bm{W}^*$.}
			\STATE{Initialize $\bm{\Phi}^{(s)}$, $\hbar^{(s)}$, and set the SCA iteration index $s=0$.}
			\REPEAT
			\STATE{Update $\bm{\Phi}^{(s+1)}$ by solving \eqref{P8-32}.}
            \STATE{Update $\hbar^{(s+1)}=\vartheta \hbar^{(s)}$.}
			\STATE{$s=s+1$.}
			\UNTIL{the increases of the objective function value of \eqref{P8-32} is less than $\epsilon_1$.}
			\STATE{{Obtain $\bm{\varphi}^{*}$ by applying the SVD of $\bm{\Phi}^{*}$, compute $\bm{\Theta}^{*}=\text{diag}\{(\bm{\varphi}^{*})^H\}$, and let $\bm{\Theta}^{(\nu)}=\bm{\Theta}^{*}$.}}
			\STATE{Calculate $\mathcal{X}^{(\nu)}=R_{sec}^{(c)}(\{\bm{w}_{m,k}^{(\nu)}\},\bm{\Theta}^{(\nu)})$ .}
			\STATE{$\nu=\nu+1$.}
			\UNTIL{ {$\mathcal{X}^{(\nu )} -\mathcal{X}^{(\nu-1)}\le \epsilon$.}}
			\STATE{Calculate the maximum objective value $\tilde{\chi}(\{\bm{t}_{m,n}\})=\mathcal{X}^{(\nu)}$ for given $\{ \bm{t}_{m,n}\}$.}
			\RETURN{$\bm{W}^*$, $\bm{\Theta}^*$, and $\tilde{\chi}(\{\bm{t}_{m,n}\})$.}
		\end{algorithmic}
	\end{algorithm}
	\begin{algorithm}
		\caption{ The DEO-based two-layer iterative framework for solving \eqref{P2}.}
		\label{Algorithm2}
		\begin{algorithmic}[1]  
			\STATE {Initialize population $\mathcal{P}_\text{DEO}^{(p)}$, the mutation scaling factor $F$, the crossover probability $P_R$, the maximum iteration number $B_2$, and set the initial iteration index $p=0$.}
			\STATE{Calculate the fitness value of the current MA positions (i.e., $\mathcal{F}(\mathcal{P}_\text{DEO}^{(p)})$) based on \eqref{Fit}, \eqref{SDEO}, and Algorithm \ref{Algorithm1}.} 
			
			\WHILE{$p \leq P$}
			\FOR{$\bar{n}=1:MN$}
			\STATE{Perform the mutation and crossover operations on the $\mathcal{P}_\text{DEO}^{(p)}$ according to \eqref{cross} and \eqref{mutation} to obtain $\mathcal{P}_\text{DEO,off}^{(p)}$.}
			\STATE{Replace a random particle in $\mathcal{P}_\text{DEO}^{(p)}$ with the $\bar{n}$-th particle of $\mathcal{P}_\text{DEO,off}^{(p)}$ to obtain $\mathcal{P}_{\text{DEO,new},\bar{n}}^{(p)}$.}
			\STATE{Calculate the fitness value of $\mathcal{P}_{\text{DEO,new},\bar{n}}^{(p)}$ based on \eqref{Fit}, \eqref{SDEO}, and Algorithm \ref{Algorithm1}, i.e.,  $\mathcal{F}(\mathcal{P}_{\text{DEO,new},\bar{n}}^{(p)})$.}
			\IF{$\mathcal{F}(\mathcal{P}_{\text{DEO,new},\bar{n}}^{(p)}) > \mathcal{F}(\mathcal{P}_\text{DEO}^{(p)})$}
			\STATE{Update $\mathcal{P}_\text{DEO}^{(p)}= \mathcal{P}_{\text{DEO,new},\bar{n}}^{(p)}$.}
			\ELSE{}
			\STATE{Update $\mathcal{P}_\text{DEO}^{(p)}=\mathcal{P}_\text{DEO}^{(p)}$.}
			\ENDIF			
			\ENDFOR
			\STATE{Update $\mathcal{P}_\text{DEO}^{(p+1)}=\mathcal{P}_\text{DEO}^{(p)}$.}
			\ENDWHILE
			\STATE{Obtain the optimal {MA positions} $\{\bm{t}_{m,n}^*\}=\mathcal{P}_\text{DEO}^{(P)}$.}
			
			\RETURN{ $\{\bm{t}_{m,n}^*\}$, $\bm{\Theta}^*$, and {$\bm{W}^*$.}}
			
		\end{algorithmic}
	\end{algorithm}
	\subsection{The Outer Layer}
	\label{outer}
	In this {subsection,} we optimize the {MA positions} based on the values of $\bm{W}_k^*$ and $\bm{\Theta}^*$ obtained in Section \ref{inner}. The sub-problem to optimize the MA positions can be given by
		\begin{align}
        		\label{P9-34}
			\max_{\{\bm{t}_{m,n}\},\chi}~~~\chi
			~~~~~{\text{s. t.}}~~\eqref{15c},\eqref{15d},\eqref{15f},\eqref{15g},\eqref{16c}.
		\end{align}
	The expansive solution space renders direct optimization of \eqref{P9-34} computationally prohibitive, as exhaustive search strategies exhibit exponential complexity {in searching the optimal MA positions.} Therefore, we apply the DEO method to address \eqref{P9-34}, which is an improved version of the conventional {differential evolution (DE) method} and includes two key characterizes, i.e., a one-in-one representation and an adaptive penalty mechanism \cite{DEO}. {According to \cite{DEO}, we first} define and initialize the population
	\begin{align}
		\label{pdeo}
		\mathcal{P}_\text{DEO}^{(0)}=\{\bar{\bm{t}}_1^{(0)},\cdots,\bar{\bm{t}}_{MN}^{(0)}\},
	\end{align}
	where the $\bar{n}$-th particle $\bar{\bm{t}}_{\bar{n}}^{(0)}=[x_{\bar{n}}^{(0)},y_{\bar{n}}^{(0)},z_{\bar{n}}^{(0)}]^T$ denotes the initial position of the $\bar{n}$-th MA among all $MN$ MAs under the
	constraint \eqref{15f}, $\bar{n}\in\bar{\mathcal{N}}=\{1,\cdots,MN\}$. From \eqref{pdeo}, it can be observed that {the one-in-one representation in the DEO method regards all MA positions as a population, where each individual particle corresponding to the position coordinates of a single MA.} This representation achieves significant dimensionality reduction in the search space, leading to substantially improved computational efficiency. Then, we introduce the fitness function incorporating the adaptive penalty mechanism to assess the influence of current MA positions in the $p$-th iteration {of the DEO method,} formulated as
	\begin{align}
		\label{Fit}
		\mathcal{F}(\mathcal{P}_\text{DEO}^{(p)})=\tilde{\chi}(\mathcal{P}_\text{DEO}^{(p)})-\omega S(\mathcal{P}_\text{DEO}^{(p)})|\mathcal{B}(\mathcal{P}_\text{DEO}^{(p)})|,
	\end{align}
	where $\tilde{\chi}(\mathcal{P}_\text{DEO}^{(p)})$ represents the objective function value obtained under the current {MA positions} $\mathcal{P}_\text{DEO}^{(p)}$ through Algorithm \ref{Algorithm1}, $\omega$ denotes a large positive scaling factor, $|\mathcal{B}(\mathcal{P}_\text{DEO}^{(p)})|$ corresponds to the cardinality of $\mathcal{B}(\mathcal{P}_\text{DEO}^{(p)})$, {representing the set of the positions of all MA pairs {which} violate constraint \eqref{15f}.} Additionally, $S(\mathcal{P}_\text{DEO}^{(p)})$ is defined as the total degree of violation of constraint \eqref{15f} under the current MA positions, which can be calculated as
	\begin{align}
		\label{SDEO}	S(\mathcal{P}_\text{DEO}^{(p)})={\sum\nolimits_{(\bar{\bm{t}}_{\bar{n}_1},\bar{\bm{t}}_{\bar{n}_2})\in\mathcal{B}(\mathcal{P}_\text{DEO}^{(p)})}}D-||\bar{\bm{t}}_{\bar{n}_1}-\bar{\bm{t}}_{\bar{n}_2}||_2,
	\end{align}
	where ${\bar{n}_1},{\bar{n}_2}\in\bar{\mathcal{N}}$.
	Note that this adaptive penalty mechanism accounts for both the number of MA pairs violating the minimum distance constraint and the degree of constraint violations with the current MA positions. This design imposes more severe penalties on MA positions with a greater degree of constraint violation to accelerate algorithmic convergence by promoting targeted solution space exploration. Next, we perform  mutation and crossover operations across all dimensions of each particle in the population $\mathcal{P}_\text{DEO}^{(p)}$ to generate the offspring population.
	For the $\bar{n}$-th particle, the detailed procedure is as follows
	\begin{align}
		\label{cross}		\bm{v}_{\bar{n}}^{(p)}=\bar{\bm{t}}_{s_1}^{(p)}+F(\bar{\bm{t}}_{s_2}^{(p)}-\bar{\bm{t}}_{s_3}^{(p)}),
	\end{align}
	\begin{equation}
		\label{mutation}
		[\bm{u}_{\bar{n}}^{(p)}]_{\bar{\ell}} = \left \{
		\begin{aligned}
			&[\bm{v}_{\bar{n}}^{(p)}]_{\bar{\ell}}, ~~~~~\text{if}~~ p_c <P_R~~\text{or}~~ \bar{\ell}=\ell', \\
			&[\bar{\bm{t}}_{\bar{n}}^{(p)}]_{\bar{\ell}}, ~~~~~\text{otherwise},
		\end{aligned}
		\right.
	\end{equation}
	where $\bar{\bm{t}}_{s_1}^{(p)}$, $\bar{\bm{t}}_{s_2}^{(p)}$, and $\bar{\bm{t}}_{s_3}^{(p)}$ denote three particles randomly selected from the population $\mathcal{P}_\text{DEO}^{(p)}$, with the exclusion of $\bar{\bm{t}}_{\bar{n}}^{(p)}$, $F$ is a mutation scaling factor, {$[\hat{x}]_i$ represents the $i$-th element of $\hat{x}$,} the random variable $p_c$ follows a uniform distribution $\mathcal{U}(0,1)$, and $P_R$ represents the crossover probability. {$\ell'$  is a number randomly selected from $\{1,2,3\}$, which guarantees that $\bm{u}_{\bar{n}}^{(p)}$ contains at least one component from $\bm{v}_{\bar{n}}^{(p)}$}. $\bar{\ell} \in\{1,2,3\}$ indicates the dimension index of the particle. Then, the offspring population in the $p$-th iteration can be denoted as $\mathcal{P}_\text{DEO,off}^{(p)}=\{{\bm{u}}_1^{(p)},\cdots,{\bm{u}}_{MN}^{(p)}\}$.
	Finally, we perform the selection replacement operation to update the population $\mathcal{P}_\text{DEO}^{(p)}$, {which is given by} 
	\begin{equation}
		\label{raplace}
		\mathcal{P}_\text{DEO}^{(p)} = \left \{
		\begin{aligned}
			&\mathcal{P}_{\text{DEO,new},\bar{n}}^{(p)}, ~~~~~\text{if}~~ \mathcal{F}(\mathcal{P}_{\text{DEO,new},\bar{n}}^{(p)})>\mathcal{F}(\mathcal{P}_\text{DEO}^{(p)}), \\
			&\mathcal{P}_\text{DEO}^{(p)}, ~~~~~~~~~~~\text{otherwise},
		\end{aligned}
		\right.
	\end{equation}
	where $\mathcal{P}_{\text{DEO,new},\bar{n}}^{(p)}$ is the population obtained by replacing a randomly selected particle in $\mathcal{P}_\text{DEO}^{(p)}$ with the $\bar{n}$-th particle from $\mathcal{P}_\text{DEO,off}^{(p)}$, $\bar{n}\in\bar{\mathcal{N}}$. {After that $\bar{n}$} has iterated through all the elements in $\bar{\mathcal{N}}$, $	\mathcal{P}_\text{DEO}^{(p+1)}=	\mathcal{P}_\text{DEO}^{(p)}$. In addition, if the updated position of the particle exceeds the specified range, we set the current position component to the corresponding boundary value of the predefined range, i.e., 
	\begin{equation}
		\mathcal{\mu}_{m,n}^{(p)} = \left \{
		\begin{aligned}
			&\mu_{m,n}^{\text{min}}~~~~~\text{if}~~ \mathcal{\mu}_{m,n}^{(p)} < \mathcal{\mu}_{m,n}^{\text{min}}, \\
			&\mu_{m,n}^{\text{max}}~~~~~\text{if}~~ \mathcal{\mu}_{m,n}^{(p)} > \mathcal{\mu}_{m,n}^{\text{max}}, \\
            &\mu_{m,n}^{(p)}~~~~~\text{otherwise}, \\
		\end{aligned}
		\right.
	\end{equation}
    where $\mu=\{x,y,z\}$, $\mu_{m,n}^{\text{max}}= \mu_m + \frac{A}{2}$, $\mu_{m,n}^{\text{min}} = \mu_m - \frac{A}{2}$, $m \in \mathcal{M}$, and $n\in \mathcal{N}$.

	\subsection{Summarization and Analysis of DEO-based Two-layer Iterative Framework}
	\label{conv1}
	{Algorithm \ref{Algorithm2} summarizes the DEO-based two-layer iterative framework for solving \eqref{P2}.

	{We then analyze the computational complexity of Algorithm \ref{Algorithm2}. First, we anlayze the computational complexity of Algorithm \ref{Algorithm1}, which is determined by the complexities of solving \eqref{P4-22} and \eqref{P8-32}, which are given by $\mathcal{O}(O_1)$ and $\mathcal{O}(O_2)$, respectively, where $O_1=\log(\frac{1}{\epsilon_1})((M+N_p(N_e+N_s+2)+N_s)M^3N^3+(MN_p(N_e+N_s+2)+N_s)^2M^2N^2+(M+N_p(N_e+N_s+2)+N_s)^3)$ and $O_2=\log(\frac{1}{\epsilon_1})((N_{\varepsilon}+N_p(N_e+N_s)+1)N_{\varepsilon}^3+(N_{\varepsilon}+N_p(N_e+N_s)+1)^2N_{\varepsilon}^2+(N_{\varepsilon}+N_p(N_e+N_s)+1)^3)$ \cite{Globally}.
    Thus, the overall computational complexity of Algorithm \ref{Algorithm1} is $\mathcal{O}(B_1(O_1+O_2))$, 
    where $B_1$ denotes the number of iterations needed for the AO framework to execute. The computational complexity of the DEO method is $\mathcal{O}(MNB_2)$ \cite{DEO}, where $B_2$ is the maximum iteration number. In summary, the overall computational complexity of Algorithm \ref{Algorithm2} can be expressed as 
    $\mathcal{J}_{DEO}=\mathcal{O}(MNB_2B_1(O_1+O_2))$.

	\section{Problem Formulation and Proposed Algorithm for {Discrete Position Case}}
	\label{discrete}
		In this section, we consider the discrete position case, where the MA positions can only be chosen from the set of quantized candidate positions. For this case, the maximization of the minimum secrecy rate of the primary transmission is given as 
	\begin{subequations}
		\label{P42}
		\begin{align}
			&\max_{\{\bm{w}_{m,k}\},\bm{\Theta},\{\bm{C}_m\},\chi}~~~~~~~ \chi \\
			&~~~{\text{s. t.}}~~\eqref{15b},\eqref{15e},\\
			\label{42c1}
			&~~~~~~~~{R_{s,a,k}^{(d)} \geq R_{th1},k\in\mathcal{N}_p,a\in\mathcal{N}_s,}\\
			\label{42c2}
			&~~~~~~~~{R_{c,a}^{(d)} \geq R_{th2},a\in\mathcal{N}_s,}\\
			\label{42c3}
			&~~~~~~~~{R_{p,k,k}^{(d)}-R_{e,b,k}^{(d)}\geq\chi,k\in\mathcal{N}_p,b\in\mathcal{N}_e,}\\
			\label{42c}
			&~~~~~~~~\bm{c}_{m,n,q} \in\{0,1\}, m\in\mathcal{M}, n\in\mathcal{N}, q\in\mathcal{Q},\\
			\label{42d}
			&~~~~~~~~{\sum\nolimits_{q=1}^Q} \bm{c}_{m,n,q} = 1,  m\in\mathcal{M}, n\in\mathcal{N},\\
			\label{42e}
			&~~~~~~~~{\sum\nolimits_{n=1}^N} \bm{c}_{m,n,q} \leq 1, m\in\mathcal{M}, q\in\mathcal{Q},\\
			\label{42f}
			&~~~~~~~~\bm{c}_{m,n_1}^T\bm{D}^m\bm{c}_{m,n_2}\geq D, n_1,n_2\in\mathcal{N}, n_1\neq n_2, m\in\mathcal{M},
		\end{align}
	\end{subequations}
	where \eqref{42c} represents that {each element of} the binary selection matrix $\bm{C}_m$ can only {be either 0 or 1.} \eqref{42d} and \eqref{42e} show that each MA can occupy only one discrete candidate position, and each position can be assigned to at most one MA.
	\eqref{42f} is the minimum distance constraint between any pair of MAs  to avoid coupling effects, in which $\bm{D}^m \in\mathbb{C}^{Q\times Q}$ denotes the distance matrix of the $m$-th AP, and the element at the $q_1$-th row and $q_2$-th column of $\bm{D}^m$, denoted by $\bm{D}_{q_1,q_2}^m$, represents the distance between the $q_1$-th and $q_2$-th discrete positions in $\bm{\mathcal{P}}_{m}$. Note that \eqref{P42} is a mixed integer programming problem, which is challenging to solve. To deal with this challenge, we propose two efficient methods. The first method is the extended DEO-based two-layer iterative framework, similar to that shown in Section \ref{two-layer}, to obtain the solution to \eqref{P42} with satisfactory accuracy. The second method is the AO iterative framework, in which the SCA, SDR, and penalty methods are used to find the discrete MA positions with low complexity. The details of the two proposed methods are described in the following.
	\subsection{Extended DEO-based Two-layer Iterative Framework}
	\label{DEO2}
	{Similar to Section \ref{two-layer}, the extended DEO-based two-layer iterative framework is comprised of the inner layer and the outer layer. In the inner layer, we follow the same procedure as shown in Section \ref{inner}  to optimize $\{\bm{w}_{m,k}\}$, $\bm{\Theta}$, and $\chi$ with $\{\bm{C}_m\}$ being fixed. In the outer layer, we {extend} the DEO to find the optimal discrete MA positions. 
    For this discrete position case, we first initialize the population as 
        \begin{align}
        \label{PiniDisDEO}
        \mathcal{P}_\text{DEO}^{(0)}=\{\bm{C}_{[1]}^{(0)},\cdots,\bm{C}_{[MN]}^{(0)}\}, 
        \end{align}
        where the $\bar{n}$-th particle, represented by 
        $\bm{C}_{[\bar{n}]}^{(0)}$, corresponds to the $\bar{n}$-th column of $\bm{C}$, $\bar{n}\in\bar{\mathcal{N}}=\{1,\cdots,MN\}$, and 
		$\bm{C}$ is formulated as 
		\begin{equation}
			\label{C}
			\bm{C} = 
			\begin{bmatrix}
				\bm{C}_1 & \bm{0}_{Q \times N} & \cdots & \bm{0}_{Q \times N} \\
				\bm{0}_{Q \times N} & \bm{C}_2 & \cdots & \bm{0}_{Q \times N} \\
				\vdots & \vdots & \ddots & \vdots \\
				\bm{0}_{Q \times N} & \bm{0}_{Q \times N} & \cdots & \bm{C}_M
			\end{bmatrix}
			\in \mathbb{C}^{MQ\times MN},
		\end{equation}
		where $\bm{0}_{Q_1 \times Q_2}$ denotes a $Q_1 \times Q_2$ zero matrix. 
        
     For the $p$-th iteration, we perform mutation and crossover operations on all dimensions (i.e., $MQ$) of each particle in $ \mathcal{P}_\text{DEO}^{(p)}$ according to \eqref{cross} and \eqref{mutation} to obtain $\mathcal{P}_\text{DEO,off}^{(p)}$.
        Due to the presence of constraints \eqref{42c}, \eqref{42d}, and \eqref{42e},
        it is necessary to map and determine each value of every particle in $\mathcal{P}_\text{DEO,off}^{(p)}$ to ensure that its value is either 0 or 1, which is given by \footnote{The performance of the extended DEO method  by introducing the mapping and determination operations will be verified in Section \ref{PerEva}.}
	\begin{align}
		\label{map}
		&\mathcal{P}_\text{DEO,off,prob}^{(p)}=\frac{1}{1+e^{-\mathcal{P}_\text{DEO,off}^{(p)}}},\\
		\label{dete}
		&\mathcal{P}_\text{DEO,off,prob}^{(p)}\overset{\text{determine}}{\longrightarrow}\mathcal{P}_\text{DEO,off}^{(p)}.
	\end{align}
	The purpose of \eqref{map} is to map each element of all columns in $\mathcal{P}_\text{DEO,off}^{(p)}$ to the probability of being selected as 1, resulting in $\mathcal{P}_\text{DEO,off,prob}^{(p)}$, where each value lies within the range $[0,1]$.
	In the determination operation defined in \eqref{dete}, we first identify the maximum value in the first column of $\mathcal{P}_\text{DEO,off,prob}^{(p)}$, record its row index $r_1^{max}$ in a defined set $r^{max}$, which is given by $r^{max}=\emptyset \cup \{r_1^{max}\}$. Then, we set the maximum value to 1, with all other values in the same column being 0. 
   When processing the $\hat{j}$-th column, $\hat{j}\in\{2,\cdots,MN\}$, the values corresponding to the row indices in $r^{max}$ are first set to negative infinity to exclude them from selection, and the row index of the maximum value in the remaining rows is then recorded as $r_{\hat{j}}^{max}$, and update $r^{max}$ as $r^{max} \cup \{r_{\hat{j}}^{max}\}$. Then, we set the value of the $r_{\hat{j}}^{max}$-th row and the $\hat{j}$-th column to 1 and the other values in the  $\hat{j}$-th column to 0.
  Thus, \eqref{dete} ensures that \eqref{42c} - \eqref{42e} are satisfied.
   
 Similar to Section \ref{outer}, $\mathcal{P}_{\text{DEO,new},\bar{n}}^{(p)}$ is formed by replacing a randomly selected column of $\mathcal{P}_{\text{DEO}}^{(p)}$ 
	with the $\bar{n}$-th column of $\mathcal{P}_{\text{DEO,off}}^{(p)}$, where $\bar{n}\in\bar{\mathcal{N}}$. 
    It is worth noting that $\mathcal{P}_{\text{DEO,new},\bar{n}}^{(p)}$ also satisfies constraints\eqref{42c} and \eqref{42e}.
    To further guarantee that $\mathcal{P}_{\text{DEO,new},\bar{n}}^{(p)}$ satisfies constraint \eqref{42d}, we need to examine $\mathcal{P}_{\text{DEO,new},\bar{n}}^{(p)}$.
    Specifically, for $\bar{j}\in\bar{\mathcal{N}}$, if the row index of the element with value 1 in the $\bar{j}$-th column of $\mathcal{P}_{\text{DEO,new},\bar{n}}^{(p)}$
	does not belong to the set $\mathcal{R}_{\bar{j}}$, \eqref{42d} is violated and $\mathcal{P}_{\text{DEO,new},\bar{n}}^{(p)}$ must be discarded, where $\mathcal{R}_{\bar{j}}$=\{$1+(m_{\bar{j}}-1)Q$, $2+(m_{\bar{j}}-1)Q$,$\cdots$,$m_{\bar{j}}Q$\}, $m_{\bar{j}}=\lfloor \frac{\bar{j}}{N+1}\rfloor +1$, and $\lfloor a \rfloor$ denotes the floor of $a$.
    
    Finally, $\mathcal{P}_\text{DEO}^{(p)}$ is updated according to \eqref{raplace}, for which the method for calculating the fitness value is given by \eqref{Fit}.
    The other steps for the DEO algorithm can refer to Section \ref{outer}.

	\subsection{AO Iterative Framework}
	\label{AO2}
	To further reduce the  computational complexity caused by the DEO-based two-layer iterative framework, we then propose an AO framework to solve problem \eqref{P42}. Specifically, problem \eqref{P42} is decomposed into three sub-problems, which are solved sequentially to obtain an approximately optimal solution. 
	The first two sub-problems are with respect to the transmit beamforming (i.e., $\{\bm{w}_{m,k}\}$, $m\in\mathcal{M}$, $k\in\mathcal{N}_p$) and the RIS phase shift matrix (i.e., $\bm{\Theta}$), respectively, whose solving method and steps are consistent with those presented in Section \ref{inner}. Thus, our focus is on optimizing $\{\bm{C}_m\}$ given optimized $\{\bm{w}_{m,k}^*\}$ and $\bm{\Theta}^*$. In particular, the third sub-problem with $\{\bm{C}_m\}$ is formulated as 
			\begin{align}
            			\label{P47}
				\max_{\{\bm{C}_m\},\chi}~~~ \chi 
				~~~{\text{s. t.}}~~\eqref{42c1}-\eqref{42f}.
			\end{align}
    To address the non-convexity of constraints \eqref{42c1}-\eqref{42c3}, auxiliary variables are introduced.
	First, let
	$\widetilde{\bm{C}}=[\bm{c}_{1,1};\cdots;\bm{c}_{1,N};\cdots;\bm{c}_{M,1};\cdots;\bm{c}_{M,N}]\in\mathbb{C}^{MNQ\times 1}$ 
	and  $\widehat{\bm{C}}=\widetilde{\bm{C}}\widetilde{\bm{C}}^H\in \mathbb{C}^{MNQ\times MNQ}$, where $\text{Rank}(\widehat{\bm{C}})=1$, and {$\widehat{\bm{C}}\succeq 0$.} Then, we introduce $\widetilde{\bm{G}}_{\xi,\zeta}\in\mathbb{C}^{MNQ\times MN}$, $\widetilde{\bm{G}}_{m,\xi,\zeta}\in\mathbb{C}^{NQ\times N}$, $\bm{F}_{\xi,\zeta}\in\mathbb{C}^{MNQ\times MN}$, and $\bm{F}_{m,\xi,\zeta}\in\mathbb{C}^{NQ\times N}$, given by 	
	\begin{equation}
		\label{Gxi}
		\widetilde{\bm{G}}_{\xi,\zeta} = 
		\begin{bmatrix}
			\widetilde{\bm{G}}_{1,\xi,\zeta} & \bm{0}_{NQ \times N} & \cdots & \bm{0}_{NQ \times N} \\
			\bm{0}_{NQ \times N} & \widetilde{\bm{G}}_{2,\xi,\zeta} & \cdots & \bm{0}_{NQ \times N} \\
			\vdots & \vdots & \ddots & \vdots \\
			\bm{0}_{NQ \times N} & \bm{0}_{NQ \times N} & \cdots & \widetilde{\bm{G}}_{M,\xi,\zeta}
		\end{bmatrix},~~~~~~~~~
	\end{equation} 
	\begin{equation}
		\label{Gxim}
		\widetilde{\bm{G}}_{m,\xi,\zeta} = 
		\begin{bmatrix}
			(\widehat{\bm{g}}_{m,\xi,\zeta}^H)^T & \bm{0}_{Q \times 1} & \cdots & \bm{0}_{Q \times 1} \\
			\bm{0}_{Q \times 1} & (\widehat{\bm{g}}_{m,\xi,\zeta}^H)^T & \cdots & \bm{0}_{Q \times 1} \\
			\vdots & \vdots & \ddots & \vdots \\
			\bm{0}_{Q \times 1} & \bm{0}_{Q \times 1} & \cdots & (\widehat{\bm{g}}_{m,\xi,\zeta}^H)^T
		\end{bmatrix},
	\end{equation}
	\begin{equation}
		\label{Fxi}
		\bm{F}_{\xi,\zeta}= 
		\begin{bmatrix}
			\bm{F}_{1,\xi,\zeta} & \bm{0}_{NQ \times N} & \cdots & \bm{0}_{NQ \times N} \\
			\bm{0}_{NQ \times N} & \bm{F}_{2,\xi,\zeta} & \cdots & \bm{0}_{NQ \times N} \\
			\vdots & \vdots & \ddots & \vdots \\
			\bm{0}_{NQ \times N} & \bm{0}_{NQ \times N} & \cdots & \bm{F}_{M,\xi,\zeta}
		\end{bmatrix},~~~~~~~~~
	\end{equation} 
		\begin{equation}
			\label{Fxim}
			\bm{F}_{m,\xi,\zeta} = 
			\begin{bmatrix}
				\widetilde{\bm{h}}_{m,\xi,\zeta}^T & \bm{0}_{Q \times 1} & \cdots & \bm{0}_{Q \times 1} \\
				\bm{0}_{Q \times 1} & \widetilde{\bm{h}}_{m,\xi,\zeta}^T & \cdots & \bm{0}_{Q \times 1} \\
				\vdots & \vdots & \ddots & \vdots \\
				\bm{0}_{Q \times 1} & \bm{0}_{Q \times 1} & \cdots & \widetilde{\bm{h}}_{m,\xi,\zeta}^T
			\end{bmatrix},~~~~~~~~~~~
		\end{equation}
		where $\widetilde{\bm{h}}_{m,\xi,\zeta}=\bm{q}_{\xi,\zeta}^H\bm{\Theta}\widehat{\bm{F}}_m\in\mathbb{C}^{1\times Q}$, $m\in\mathcal{M}$, $\xi\in\{p,e,s\}$, and $\zeta\in\mathcal{N}_{\xi}$.
		In addition, we define $\widetilde{\bm{h}}_{\xi,\zeta}=[\widetilde{\bm{h}}_{1,\xi,\zeta},\cdots,\widetilde{\bm{h}}_{M,\xi,\zeta}]^H\in\mathbb{C}^{MQ\times1}$ and  $\widetilde{\bm{g}}_{\xi,\zeta}=[\widehat{\bm{g}}_{1,\xi,\zeta}^H,\cdots,\widehat{\bm{g}}_{M,\xi,\zeta}^H]^H\in\mathbb{C}^{MQ\times1}$.
	{Therefore, $\bm{h}_{\xi,\zeta}^{(d)}$ and $\bm{g}_{\xi,\zeta}^{(d)}$ described in Section \ref{trassmission} can be reformulated as 
		$\bm{h}_{\xi,\zeta}^{(d)}=\bm{C}^H\widetilde{\bm{h}}_{\xi,\zeta}$ and  $\bm{g}_{\xi,\zeta}^{(d)}=\bm{C}^H\widetilde{\bm{g}}_{\xi,\zeta}$, respectively.
		Then, by introducing $\bm{\mu}_{\xi,\zeta,k}=\widetilde{\bm{G}}_{\xi,\zeta}\bm{w}_k\in\mathbb{C}^{MNQ\times 1}$, $\bm{\mu}_{f,\xi,\zeta,k}=	\bm{F}_{\xi,\zeta}\bm{w}_k\in\mathbb{C}^{MNQ\times 1}$, $\bm{U}_{\xi,\zeta,k}=\bm{\mu}_{\xi,\zeta,k}\bm{\mu}_{\xi,\zeta,k}^H\in\mathbb{C}^{MNQ\times MNQ}$, and $\bm{U}_{f,\xi,\zeta,k}=\bm{\mu}_{f,\xi,\zeta,k}\bm{\mu}_{f,\xi,\zeta,k}^H\in\mathbb{C}^{MNQ\times MNQ}$, where $k\in\mathcal{N}_p$, $\zeta\in\mathcal{N}_{\xi}$, \eqref{42c1} and \eqref{42c2} can be reformulated as} 
	\begin{equation}
		\label{52}
		\left \{
		\begin{aligned}
			&\frac{\text{Tr}(\widehat{\bm{C}}\bm{U}_{s,a,k})}{\sum_{k' = k + 1}^{N_p} \text{Tr}(\widehat{\bm{C}}\bm{U}_{s,a,k'}) + \alpha \sum_{k' = 1}^{N_p} \text{Tr}(\widehat{\bm{C}}\bm{U}_{f,s,a,k}) + \sigma_{s}^{2}}\\
				&~~~~~~~~~~~~~~~\geq 2^{R_{{th1}}} - 1, k \in \{1, \dots, N_p - 1\}, a\in\mathcal{N}_s, \\
			&\frac{\text{Tr}(\widehat{\bm{C}}\bm{U}_{s,a,N_p})}{\alpha \sum_{k' = 1}^{N_p} \text{Tr}(\widehat{\bm{C}}\bm{U}_{f,s,a,k'}) + \sigma_{s}^{2}}\geq 2^{R_{{th1}}} - 1, a\in\mathcal{N}_s,
		\end{aligned}
		\right.
	\end{equation}
	\begin{align}
		\label{53}
		{\alpha \sum\nolimits_{k' = 1}^{N_p} \text{Tr}(\widehat{\bm{C}}\bm{U}_{f,s,a,k'})}\geq \beta^* {\sigma_s^2}, a\in\mathcal{N}_s.
	\end{align}
    Similar to the approach used for \eqref{16c} in Section \eqref{inner}, we rewrite the {left-hand side of \eqref{42c3} as $h_1-h_2-h_3+h_4$,} where
	\begin{align}
    &h_1=\log_{2}(\sum_{k'=1}^{N_p}\text{Tr}(\widehat{\bm{C}}\bm{U}_{p,k,k'})+\alpha\sum_{k'=1}^{N_p}\text{Tr}(\widehat{\bm{C}}\bm{U}_{f,p,k,k'})+\sigma_p^2),\notag\\ 
    &h_2=\log_{2}(\sum_{k'\neq k}^{N_p}\text{Tr}(\widehat{\bm{C}}\bm{U}_{p,k,k'})+\alpha\sum_{k'=1}^{N_p}\text{Tr}(\widehat{\bm{C}}\bm{U}_{f,p,k,k'})+\sigma_p^2),\notag\\ 
    &h_3=\log_{2}(\text{Tr}(\widehat{\bm{C}}\bm{U}_{e,b,k})+\alpha\sum\nolimits_{k'=1}^{N_p}\text{Tr}(\widehat{\bm{C}}\bm{U}_{f,e,b,k'})+\sigma_e^2),\notag\\ 
    &h_4=\log_{2}(\alpha\sum\nolimits_{k'=1}^{N_p}\text{Tr}(\widehat{\bm{C}}\bm{U}_{f,e,b,k'})+\sigma_e^2), k\in\mathcal{N}_p,\ b\in\mathcal{N}_e.\notag
	\end{align}    
	At any feasible point $\widehat{\bm{C}}^{(\tau)}$, the upper bounds of $h_2$ and $h_3$
	can be obtained using the first-order Taylor expansion, which are given by \eqref{49} and \eqref{50} shown at the top of this page.
\newcounter{TempEqCnt} %
\setcounter{TempEqCnt}{\value{equation}} %
\setcounter{equation}{55} %
\begin{figure*}[ht] %
  \begin{equation}
    \begin{aligned}
    \label{49}
	h_2\leq\bar{h}_2\triangleq 
	\log_{2}(\sum\nolimits_{k'\neq k}^{N_p}&\text{Tr}(\widehat{\bm{C}}^{(\tau)}\bm{U}_{p,k,k'})+\alpha\sum\nolimits_{k'=1}^{N_p}\text{Tr}(\widehat{\bm{C}}^{(\tau)}\bm{U}_{f,p,k,k'})+\sigma_p^2)\\
    &+\frac{\sum\nolimits_{k'\neq k}^{N_p}\text{Tr}((\widehat{\bm{C}}-\widehat{\bm{C}}^{(\tau)})\bm{U}_{p,k,k'})+\alpha\sum_{k'=1}^{N_p}\text{Tr}((\widehat{\bm{C}}-\widehat{\bm{C}}^{(\tau)})\bm{U}_{f,p,k,k'})}{(\sum_{k'\neq k}^{N_p}\text{Tr}(\widehat{\bm{C}}^{(\tau)}\bm{U}_{p,k,k'})+\alpha\sum_{k'=1}^{N_p}\text{Tr}(\widehat{\bm{C}}^{(\tau)}\bm{U}_{f,p,k,k'})+\sigma_p^2)\ln2},\\
  \end{aligned}
  \end{equation}
    \vspace{-0.6cm}
\end{figure*}
\setcounter{TempEqCnt}{\value{equation}} 
\setcounter{equation}{56} 
\begin{figure*}[ht] 
  \begin{equation}
    \begin{aligned}
    \label{50}    
	h_3\leq\bar{h}_3\triangleq \log_{2}(\text{Tr}(\widehat{\bm{C}}^{(\tau)}&\bm{U}_{e,b,k})+\alpha\sum\nolimits_{k'=1}^{N_p}\text{Tr}(\widehat{\bm{C}}^{(\tau)}\bm{U}_{f,e,b,k'})+\sigma_e^2)~~~~~~~~~~~~~~~~~~~~~~~~~~~~~~~\\
    &+\frac{\text{Tr}((\widehat{\bm{C}}-\widehat{\bm{C}}^{(\tau)})\bm{U}_{e,b,k})+\alpha\sum\nolimits_{k'=1}^{N_p}\text{Tr}((\widehat{\bm{C}}-\widehat{\bm{C}}^{(\tau)})\bm{U}_{f,e,b,k'})}{(\text{Tr}(\widehat{\bm{C}}^{(\tau)}\bm{U}_{e,b,k})+\alpha\sum\nolimits_{k'=1}^{N_p}\text{Tr}(\widehat{\bm{C}}^{(\tau)}\bm{U}_{f,e,b,k'})+\sigma_e^2)\ln2}.
  \end{aligned}
  \end{equation}
  \hrulefill 
    \vspace{-0.6cm}
\end{figure*}

Based on the mathematical operations above, \eqref{42c3} can be rewritten as 
	\begin{align}
		\label{54}
		h_1-\bar{h}_2-\bar{h}_3+h_4\geq \chi.
	\end{align}
	Taking into account the structure of $\widehat{\bm{C}}$, \eqref{42d} and \eqref{42e} {can be respectively rewritten as} 
	\begin{align}
		\label{55}
		&\sum\widehat{\bm{C}}(1+(\bar{a}-1)Q:\bar{a}Q,1+(\bar{b}-1)Q:\bar{b}Q)=1,\notag\\& ~~~~~~~~~~~~~~~~~~~~~~~~~~~~~~~~\bar{a},\bar{b}\in\{1,2,\cdots,MN\},\\
		\label{56}
		&{\sum\nolimits_{n=1}^{N}}\sum\widehat{\bm{C}}(:,(m-1)NQ+q+(n-1)Q) \leq MN,~~~\notag\\&~~~~~~~~~~~~~~~~~~~~~~~~~~~~~~~~m\in\mathcal{M}, q\in\mathcal{Q}.
	\end{align}
	Furthermore, to address the non-convexity of \eqref{42f}, auxiliary variables {are also} introduced. Let $\widetilde{\bm{c}}_m=[\bm{c}_{m,1}^T,\cdots,\bm{c}_{m,N}^T]^T\in\mathbb{C}^{QN\times 1}$, $\widetilde{\bm{{I}}}_n=[\bm{0}_{Q\times (n-1)Q},\bm{{I}}_{Q\times Q},\bm{0}_{Q\times (N-n)Q}]\in\mathbb{C}^{Q\times QN}$, and $\bm{D}_{n_1,n_2}^m=(\widetilde{\bm{{I}}}_{n_1})^T\bm{D}^m\widetilde{\bm{{I}}}_{n_2}$, where $\bm{{I}}_Q$ denotes the $Q$-dimensional identity matrix, $m\in\mathcal{M}$, $n_1\neq n_2$, and $n,n_1,n_2\in\mathcal{N}$. 
	{Following the derivation details shown in \cite{Globally}, we can transform \eqref{42f} into \eqref{57},} expressed as  
	\begin{align}
		\label{57}
		{\widetilde{\bm{c}}_m^T(({-\bm{D}_{n_1,n_2}^m-\bm{D}_{n_1,n_2}^{m,T})}/{2}+\widetilde{\alpha}\bm{{I}}_{QN})\widetilde{\bm{c}}_m-\widetilde{\alpha}N+D\leq0,}
	\end{align}
	where $\widetilde{\alpha}\geq \max\{ \bar{\lambda}(\frac{\bm{D}_{n_1,n_2}^m+(\bm{D}_{n_1,n_2}^m)^T}{2})\}$ {corresponding} to the largest eigenvalue of $\frac{\bm{D}_{n_1,n_2}^m+(\bm{D}_{n_1,n_2}^m)^T}{2}$. Subsequently, using the Schur complement lemma \cite{Schur}, \eqref{57} can be restated as 
	\begin{equation}
		\label{schur}
		\begin{bmatrix}
			({({-\bm{D}_{n_1,n_2}^m-\bm{D}_{n_1,n_2}^{m,T})}/{2}+\widetilde{\alpha}\bm{{I}}_{QN}})^{-1} & \widetilde{\bm{c}}_m  \\
			\widetilde{\bm{c}}_m^T & \widetilde{\alpha}N-D 
		\end{bmatrix}\succeq  0,
	\end{equation}
	where $(\mathcal{\bm{A}})^{-1}$ denotes the inverse of $\mathcal{\bm{A}}$, $\widetilde{\bm{c}}_m=\bm{\breve{C}}(1+(m-1)QN:mQN,:)$, and $\bm{\breve{C}}=\left[\sqrt{\widehat{\bm{C}}(1,1)};\cdots;\sqrt{\widehat{\bm{C}}(MNQ,MNQ)} \right]\in\mathbb{C}^{MNQ\times 1}$. To circumvent the non-convexity induced by  square-root operations, we introduce a penalty term $\sum\nolimits_{i=1}^{MNQ}(\widehat{\bm{C}}(i,i)-{\bm{\breve{C}}(i)^2})^2$ into the objective function. A first-order Taylor expansion is then performed at any feasible point $\bm{\breve{C}}^{(\tau)}$, yielding $\digamma(\widehat{\bm{C}},\bm{\breve{C}},\bm{\breve{C}}^{(\tau)})=\sum\nolimits_{i=1}^{MNQ}(\widehat{\bm{C}}(i,i)-{\bm{\breve{C}}^{(\tau)}(i)^2}-2{\bm{\breve{C}}^{(\tau)}(i)}({\bm{\breve{C}}(i)}-{\bm{\breve{C}}^{(\tau)}(i)}))^2$.
    To deal with the challenges resulting from the binary integer variables defined in \eqref{42c}, we relax them to continuous variables within the interval $[0,1]$. In addition, the  penalty method described in Section \ref{RISopt} is employed to address the non-convexity of constraint  $\text{Rank}(\widehat{\bm{C}})=1$, thereby obtaining a rank-one solution as described in Section \ref{RISopt}.
	Accordingly, \eqref{P47} is transformed into \eqref{P59}, given by
	\begin{equation}
		\label{P59}
		\begin{aligned}
			&\max_{\widehat{\bm{C}}\succeq 0,\bm{\breve{C}},\chi}~ \chi-\hbar \digamma(\widehat{\bm{C}},\bm{\breve{C}},\bm{\breve{C}}^{(\tau)})-\hbar\ell(\widehat{\bm{C}},\widehat{\bm{C}}^{(\tau)})~~\\
			&~~~~~\text{s. t.}~~\eqref{52}, \eqref{53}, \eqref{54}, \eqref{55}, \eqref{56}, \eqref{schur},
		\end{aligned}
	\end{equation}
    where $\ell(\widehat{\bm{C}},\widehat{\bm{C}}^{(\tau)})$ follows the expression in \eqref{pentaylor}.
	Problem \eqref{P59} is a convex optimization problem, and its optimal solution obtained {via the CVX tool \cite{CVX}} is denoted as $\widehat{\bm{C}}^*$. Denote the eigenvector corresponding to the maximum eigenvalue of $\widehat{\bm{C}}^*$ as $\widetilde{\bm{C}}^*$, and let $\bar{\bm{C}}_m^*(:,n)=\widetilde{\bm{C}}^*((m-1)QN+(n-1)Q+1:(m-1)QN+nQ,:)$, where $n\in\mathcal{N}$ and  $m\in\mathcal{M}$. {Then, we perform the mapping and determination operations on $\{\bar{\bm{C}}_m^*\}$, similar to \eqref{map} and \eqref{dete},  to obtain the binary integer solution.}
     The detailed procedure for solving \eqref{P59} is summarized in Algorithm \ref{Algorithm3}.

		\subsection{Summarization and Analysis of Two Proposed Methods}
		\subsubsection{Extended DEO-based Two-layer Iterative Framework}
        \label{DEOanalysis}
        The computational complexity of the DEO-based two-layer algorithm is consistent with the discussion provided in Section \ref{conv1}, which is thus not reiterated here.
		\subsubsection{AO Iterative Framework}
        \label{AOanalysis}
		The computational complexity of the AO iterative algorithm is primarily determined by the three sub-problems derived from the decomposition of  \eqref{P42}. Section \ref{conv1} provides the exact expressions for the complexities associated with the first two sub-problems optimizing the transmit beamforming and RIS phase shift matrix. The complexity of the {third sub-problem, i.e., \eqref{P59},} is expressed as $\mathcal{O}(O_3)$, where $O_3=\log(\frac{1}{\epsilon_1})((M^2N^2+(N_p+1)N_s+M(Q+1)+1
        )M^3N^3Q^3+(M^2N^2+(N_p+1)N_s+M(Q+1)+1)^2M^2N^2Q^2+(  M^2N^2+(N_p+1)N_s+M(Q+1)+1)^3)$.
		Therefore, the overall computational complexity of the AO iterative algorithm for solving \eqref{P42} is $\mathcal{J}_{AO}=\mathcal{O}(B_3(O_1+O_2+O_3))$, where $B_3$ represents the number of iterations needed for the AO framework to complete, and $\epsilon_1$ is the convergence threshold.
 \subsubsection{Computational Complexity Comparison}
    $\mathcal{J}_{DEO}$ and $\mathcal{J}_{AO}$ can be reformulated as $\mathcal{J}_{DEO}=\mathcal{O}(B_1(O_1+O_2)+(MNB_2-1)B_1(O_1+O_2))$ and $\mathcal{J}_{AO}~~=\mathcal{O}(B_3(O_1+O_2)+B_3O_3)$, respectively. 
 Based on empirical observations, $(MNB_2-1)B_1$ significantly exceeds $B_3$ owing to the scaling effects of $M$ and $N$, whereas the values of $B_1$ and $B_3$ are close. Moreover, $O_1+O_2$ consistently dominates $O_3$ across typical scenarios. Thus, the inequality $\mathcal{J}_{DEO} >\mathcal{J}_{AO} $ holds.

	\begin{algorithm}
		\caption{ SCA algorithm for solving \eqref{P47}.}
		\label{Algorithm3}
		\begin{algorithmic}[1]  
			\STATE{Initialize $\widehat{\bm{C}}^{(\tau)}$ and set the SCA iteration index $\tau=0$. Set the convergence threshold as $\epsilon_1$ and  $\widetilde{\alpha}\geq \max\{ \bar{\lambda}(\frac{\bm{D}_{n_1,n_2}^m+(\bm{D}_{n_1,n_2}^m)^T}{2})\}$. }
			\REPEAT
			\STATE{Update $\widehat{\bm{C}}^{(\tau+1)}$ by solving \eqref{P59}.}
			\STATE{$\tau=\tau+1$.}
			\UNTIL{{the increases of the objective function value of \eqref{P59} is less than $\epsilon_1$.}}
			\STATE{Set $\widehat{\bm{C}}^*=\widehat{\bm{C}}^{(\tau+1)}$ .}
			\STATE{Perform the SVD on $\widehat{\bm{C}}^*$ to obtain $\widetilde{\bm{C}}^*$.}
			\STATE{{Obtain $\{\bar{\bm{C}}_m^*\}$ from $\widetilde{\bm{C}}^*$.}}
            \STATE{{Performing mapping and determination operations on $\{\bar{\bm{C}}_m^*\}$ to obtain the binary integer solution $\{\bm{C}_m^*\}$.}}
			\RETURN{$\{\bm{C}_m^*\}$.}
		\end{algorithmic}
	\end{algorithm}

\section{Numerical Results}
\label{numerical}
\subsection{Simulation Setup}
	Numerical simulations are conducted in this section to assess the performance of the proposed schemes and algorithms for continuous and discrete position cases. The parameter settings are as follows: $M=2$, $N=3$, $N_\varepsilon=25$, $N_p=2$, $N_e=2$, $N_s=2$, $P_\text{max}=40\ \text{dBm}$, $L_{t,\kappa}^m=L_{r,\kappa}^m=\bar{L}=10$, $\lambda=0.06\ \text{m}$, $D=0.5\lambda$, $A=A_1=A_2=A_3=2\lambda$, $\alpha=1$, $\sigma_{\xi}^2=-40\ \text{dBm}$, $R_{th1}=0.6\ \text{bps/Hz}$, $R_{th2}=5\ \text{bps/Hz}$, $Q=16$, $F=0.9$, $P_R=0.9$, $\epsilon=0.01$, $\epsilon_1=0.01$, $w=100$, and $B_2=200$. The location coordinates of the APs are set at $(20\ \text{m},\ 20\sqrt{3}\ \text{m},\ 10\ \text{m})$ and $(-20\ \text{m},\ 20\sqrt{3}\ \text{m},\ 10\ \text{m})$, respectively. The RIS is located at $(0\ \text{m},\ 6\ \text{m},\ 10\ \text{m})$. 
    In addition, the PUs, Eves and SUs are randomly distributed within circular areas of 2 meter radius with the center coordinates of $(0\ \text{m},\ 0\ \text{m},\ 10\ \text{m})$, $(-5\ \text{m},\ 0\ \text{m},\ 10\ \text{m})$, and $(10\ \text{m},\ 0\ \text{m},\ 10\ \text{m})$, respectively. The path-response matrices $\bm{\Sigma}_{m,\varepsilon}$ and $\bm{\Sigma}_{m,\xi,\zeta}$ are defined as $\text{diag}\{[\sigma_{m,\varepsilon}^1,\cdots,\sigma_{m,\varepsilon}^{\bar{L}}]\}$ and $\text{diag}\{[\sigma_{m,\xi,\zeta}^1,\cdots,\sigma_{m,\xi,\zeta}^{\bar{L}}]\}$, respectively, where each element follows the i.i.d. CSCG distribution $\mathcal{CN}(0, c_0 \cdot d^{-\varrho}/\bar{L})$. Here, $c_0 = -10$ dB denotes the path-loss constant, $d$ represents the distance between two nodes, and the path-loss exponent $\varrho$ is set to 2.6 for the direct link and 1.5 for the backscattering link. The elevation and azimuth AoAs/AoDs are uniformly distributed within $[-\pi/2,\ \pi/2]$. To demonstrate the superiority of the proposed schemes, we consider the following schemes in the continuous position case for comparison.
    \begin{itemize}
        \item FPA scheme: The FPAs are deployed at each AP, and  the transmit beamforming vectors at the APs and the phase shift matrix of the RIS are jointly optimized.
        \item Random passive beamforming: Based on the model shown in Section \ref{system model}, we only optimize the transmit beamforming vectors and MA positions at the APs with a random generation of the RIS phase shift matrix.
        \item Random transmit beamforming: Based on the model shown in Section \ref{system model}, we only optimize the  MA positions at the APs and the passive beamforming at the RIS with a random generation of the transmit beamforming.
    \end{itemize}
\subsection{Performance Evaluations}
\label{PerEva}
Fig. \ref{fig:SL} demonstrates the convergence of the DEO-based two-layer iterative framework (i.e., Algorithm \ref{Algorithm2}) in both continuous and discrete position cases, as well as the AO iterative framework in the discrete position case.
From Fig. \ref{fig:SL} (a), the secrecy rates attained by Algorithm \ref{Algorithm2} in both continuous and discrete cases increase with the number of iterations and eventually converge to determined values.
In addition, we can observe that compared with the discrete position case, Algorithm \ref{Algorithm2} can achieve a higher secrecy rate in the continuous position case. This is because the feasible solution range of the continuous position case is larger, so as to obtain more spatial DoFs to reconstruct the channels, which significantly improves the system performance.
As shown in Fig. \ref{fig:SL} (b), the secrecy rate obtained by the AO iterative framework is a non-decreasing function of the number of iterations, and the secrecy rate converges to a fixed value after 7 iterations, indicating that the excellent convergence performance of the AO iterative framework.

	\begin{figure}
		\centering
		\subfloat[
        {\makebox[0.1\textwidth][l]{Algorithm \ref{Algorithm2} in continuous} \\ 
         \makebox[0.1\textwidth][l]{~~~~ and discrete position cases.}
        } 
                 ]{
            \includegraphics[width=0.18\textwidth]{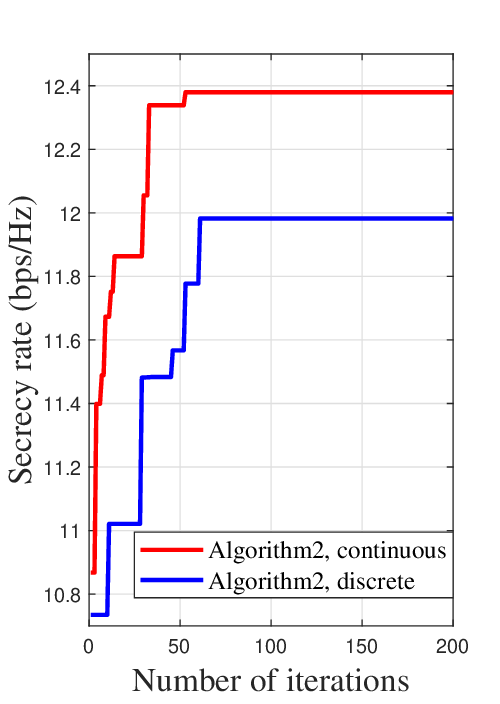}}
		~~~~~~\subfloat[
        {\makebox[0.1\textwidth][l]{The AO iterative framework} \\ 
         \makebox[0.1\textwidth][l]{~~~~ in the discrete position case.}
        }
                       ]{
			\includegraphics[width=0.18\textwidth]{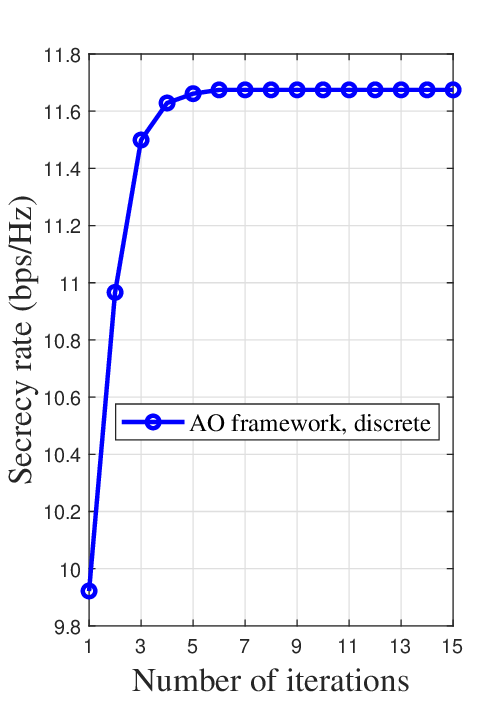}}
		 \caption{Convergence performance of the proposed algorithms. 
        }
		\label{fig:SL}
	\end{figure}

Fig. \ref{fig:P} illustrates the secrecy rate versus the maximum transmit power at each AP (i.e., $P_{\text{max}}$). As $P_{\text{max}}$ increases, the secrecy rates of all schemes increase. Among all the schemes, the proposed scheme with the DEO-based two-layer
iterative framework in the continuous position case demonstrates the best performance. Specifically, when $P_{\text{max}}=40\ \text{dBm}$, it achieves approximately $12.4\%$ performance improvement compared to the FPA scheme, confirming that the MA technology attains superior capabilities to improve system performance. Additionally, in contrast to the proposed scheme with the DEO-based two-layer
iterative framework in the discrete position case, the secrecy rate of the  proposed scheme with the DEO-based two-layer
iterative framework in the continuous position case is improved by approximately $6.9\%$, which again confirms that the performance superiority of  the continuous position case.

	\begin{figure}[t]
	\centering
	\includegraphics[width=0.4\textwidth]{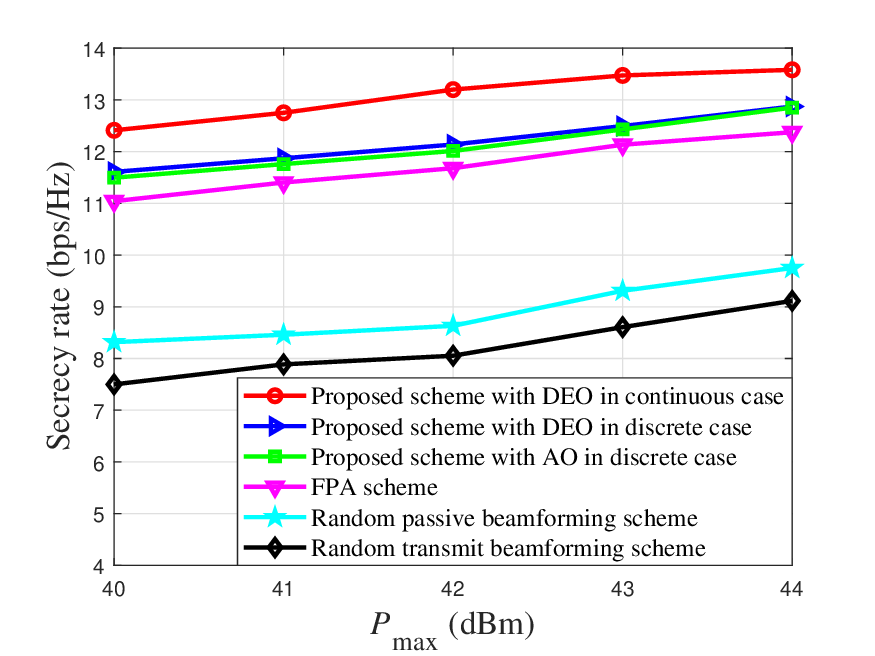} 
	\caption{Secrecy rate versus the maximum transmit power.}
	\label{fig:P}
\end{figure}

\begin{figure}[t]
	\centering
	\includegraphics[width=0.4\textwidth]{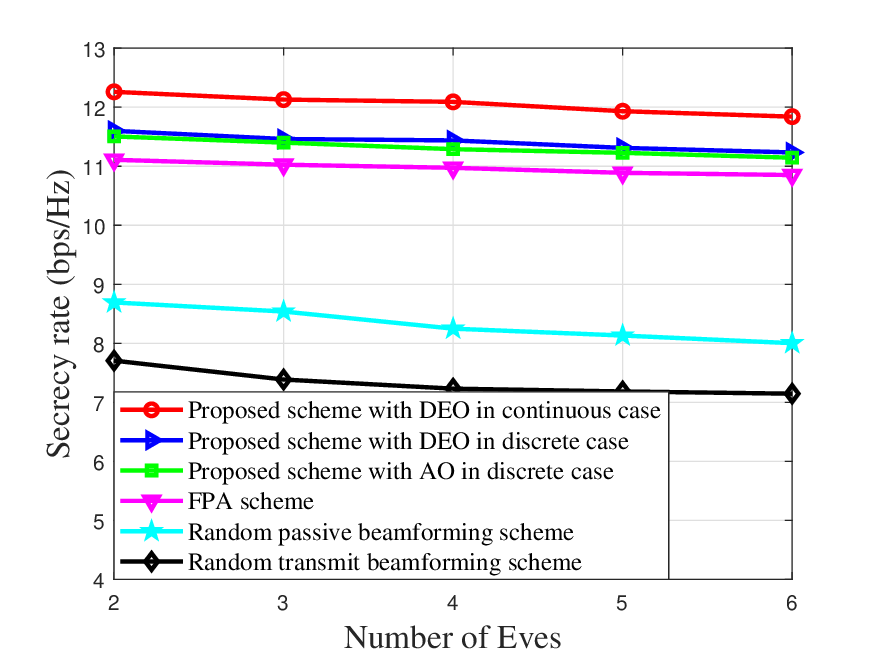} 
	\caption{Secrecy rate versus the number of Eves.}
	\label{fig:EVE}
\end{figure}

\begin{figure}[t]
	\centering
	\includegraphics[width=0.4\textwidth]{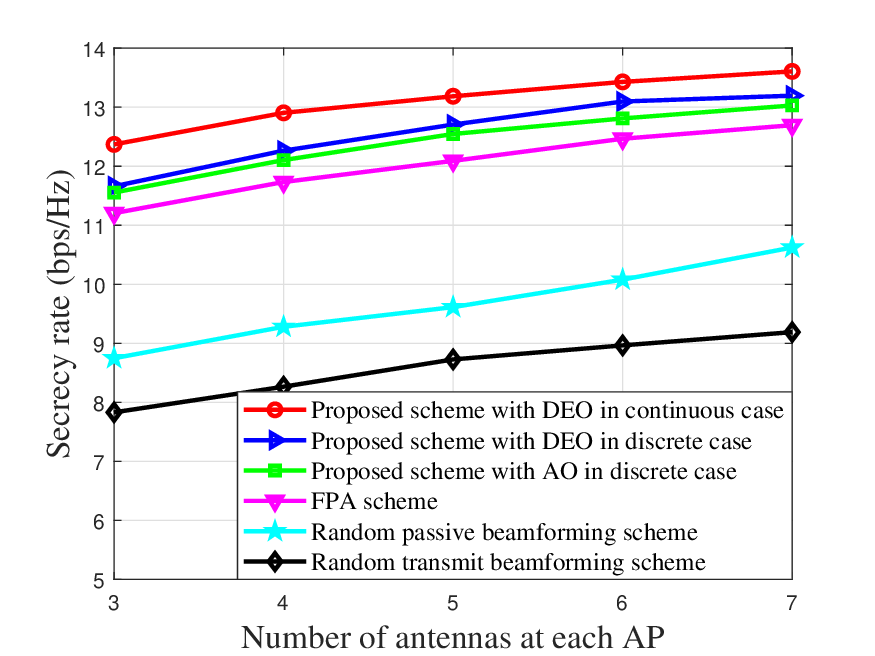} 
	\caption{Secrecy rate versus the number of antennas at each AP.}
	\label{fig:MA}
\end{figure}

The secrecy rate versus the number of Eves (i.e., $N_e$) is presented in Fig. \ref{fig:EVE}. When more Eves exist,  the confidential information transmitted from the APs to the PUs is much more likely to be intercepted illegally, leading to a lower system secrecy rate.
Moreover, the secrecy rate achieved by the proposed schemes demonstrate a significant improvement compared to the random transmit beamforming scheme and the random passive beamforming scheme. The reason  is that the two benchmark schemes may reduce the desired  signal strength at the PUs but improve that at the Eves. This also validates the effectiveness and necessity of designing transmit beamforming and passive beamforming.

Fig. \ref{fig:MA} depicts the secrecy rate versus the number of antennas at each AP (i.e., $N$). It is obvious that deploying more MAs leads to the improvement of the secrecy rate for all schemes. This can be attributed to the fact that deploying additional MAs  enables the APs to obtain more spatial diversity through the introduction of multiple independent transmission paths, making the beams transmitted from the APs to the PUs more directional  and reducing the information intercepted by the Eves. Additionally, for the discrete position case, the secrecy rate obtained by the proposed scheme with the AO iterative framework is only $1.55\%$ lower on average than that of the proposed scheme with the DEO-based two-layer iterative framework when $P_{\text{max}}=40\ \text{dBm}$, demonstrating the effectiveness of the AO iterative framework.

\begin{figure}[t]
	\centering
	\includegraphics[width=0.4\textwidth]{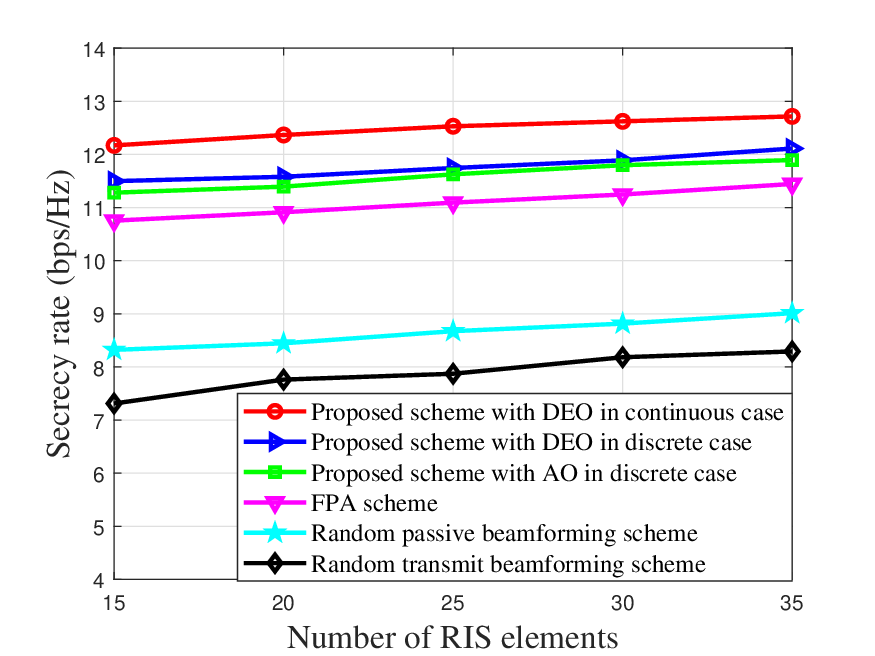} 
	\caption{Secrecy rate versus the number of RIS elements.}
	\label{fig:RIS}
\end{figure}

Fig. \ref{fig:RIS} shows the secrecy rate versus the number of RIS elements (i.e., $N_\varepsilon$). As observed, the number of RIS elements has a positive impact on the secrecy rate of the primary transmission. 
This can be attributed to two aspects. First, additional RIS elements introduce more transmission paths to assist the secondary transmission, making it easier to satisfy the QoS constraints at the SUs. Building on this, the system can prioritize allocating more resources to the primary transmission, thereby enhancing the performance of  primary transmission. Second, the RIS with more elements can impose more interference on Eves, thereby suppressing the information eavesdropping.
}

\section{Conclusion}
\label{conclusion}
In this paper, we proposed to leverage the MA technology in a cell-free RIS aided SR system to establish the secure transmission from distributed APs to PUs to against the eavesdropping from malicious Eves, and simultaneously boost the secondary transmission from the RIS to multiple SUs. For both continuous and discrete position cases,  we maximized the minimum secrecy rate of primary transmission by taking into account the QoS constraints on secondary transmission,  respectively. In particular, we proposed a DEO-based two-layer iterative framework for the continuous position case to achieve the best performance as the exploitation of continuous spatial DoFs. For the discrete position case, we first extended the DEO-based two-layer iterative framework to optimize the discrete MA positions, and then proposed an AO framework with low computational complexity  to attain a  solution with  satisfactory accuracy. Finally, we conducted comprehensive  numerical simulations to verify the superior performance of the proposed schemes and algorithms.

\end{document}